\documentclass[twocolumn,english]{IEEEtran}
\usepackage[T1]{fontenc}
\usepackage[latin9]{inputenc}
\usepackage{color}
\usepackage{array}
\usepackage{verbatim}
\usepackage{multirow}
\usepackage{amsmath}
\usepackage{amssymb}
\usepackage{graphicx}
\usepackage{esint}

\makeatletter

\providecommand{\tabularnewline}{\\}

\usepackage{babel}
\ifCLASSOPTIONcompsoc
\else
\fi
\usepackage{cite}
\usepackage{subfigure}

\usepackage{babel}

\usepackage{babel}

\makeatother

\usepackage{babel}
\begin{document}

\title{Convergence of Desynchronization Primitives in Wireless Sensor Networks:
A Stochastic Modeling Approach }

\author{Dujdow Buranapanichkit, Nikos Deligiannis, \emph{Member, IEEE,} and
\\ Yiannis Andreopoulos\emph{$^{*}$, Senior Member, IEEE}%
\thanks{$^{*}$Corresponding author. 

D. Buranapanichkit is with the Department of Electrical Engineering, Faculty of Engineering,
Prince of Songkla University, Hat Yai, Songkla, 90112 Thailand; Tel.
+66 74 287244, Fax +66 74 459395, Email: dujdow.b@psu.ac.th. 

N. Deligiannis and Y. Andreopoulos are with the Electronic and Electrical Engineering Department,
University College London, Roberts Building, Torrington Place, London,
WC1E 7JE, U.K.; Tel. +44 20 7679 7303, Fax. +44 20 7388 9325 (both
authors), Email: \{n.deligiannis, i.andreopoulos\}@ucl.ac.uk. This
work was supported in part by the EPSRC, grant EP/K033166/1. 

This paper will appear in the IEEE Trans. on Signal Processing. This is the authors' version (Nov. 11, 2014), which may slightly differ from the published version. 

Copyright (c) 2014 IEEE. Personal use of this material is permitted.
However, permission to use this material for any other purposes must
be obtained from the IEEE by sending a request to pubs-permissions@ieee.org.%
}}
\maketitle
\begin{abstract}
Desynchronization approaches in wireless sensor networks converge
to time-division multiple access (TDMA) of the shared medium without
requiring clock synchronization amongst the wireless sensors, or indeed
the presence of a central (coordinator) node. All such methods are
based on the principle of \textit{reactive listening }of periodic
``fire\textquotedblright{} or ``pulse\textquotedblright{} broadcasts:
each node updates the time of its fire message broadcasts based on
received fire messages from some of the remaining nodes sharing the
given spectrum. In this paper, we present a novel framework to estimate
the required iterations for convergence to fair TDMA scheduling. Our
estimates are fundamentally different from previous conjectures or
bounds found in the literature as, for the first time, convergence
to TDMA is defined in a \emph{stochastic sense}. Our analytic results
apply to the \textsc{Desync} algorithm and to pulse-coupled oscillator
algorithms with inhibitory coupling. The experimental evaluation via
iMote2 TinyOS nodes (based on the IEEE 802.15.4 standard) as well
as via computer simulations demonstrates that, for the vast majority
of settings, our stochastic model is within one standard deviation
from the experimentally-observed convergence iterations. The proposed
estimates are thus shown to characterize the desynchronization convergence
iterations significantly better than existing conjectures or bounds.
Therefore, they contribute towards the analytic understanding of how
a desynchronization-based system is expected to evolve from random
initial conditions to the desynchronized steady state. 
\end{abstract}


\section{Introduction}

\IEEEPARstart{E}{fficient} usage of shared spectrum in distributed
(ad-hoc) networking architectures is important for high data-rate
communications \cite{simeOne2008distributed}. This is particularly
so for wireless sensor networks (WSNs), where wasting packet transmissions
due to collisions in the medium access also means wasting battery
resources \cite{simeOne2008distributed,leidenfrost2009firefly,scaglione2010bioinspired,Degesys2007DESYNC,mutazono2011energy,yamamoto2011inter,nakano2011biologically,klinglmaye2012selforganizing}.
Desynchronization is a new WSN primitive leading to fair time-division
multiple access (TDMA) scheduling that does not require the presence
of a coordinating node \cite{scaglione2010bioinspired,Degesys2007DESYNC,yamamoto2011inter,nakano2011biologically,patel2007desync,ashkiani2012discrete,choochaisriArtificialForceField,lien2012anchored,bojic2012self}.
All desynchronization approaches are based on the principle of \textit{reactive
listening}, where nodes periodically broadcast short packets (so-called
``beacon\textquotedblright{} or ``fire\textquotedblright{} messages
\cite{pagliari2011scalable,Degesys2007DESYNC,yamamoto2011inter,nakano2011biologically,patel2007desync,ashkiani2012discrete,choochaisriArtificialForceField,lien2012anchored})
and then update their next broadcast time based on the reception of
fire messages from some of the remaining nodes. These methods make
use of a convergence interval, where nodes adjust their firing times,
and a \emph{steady-state} (SState) period. In the SState period, nodes
have converged into fair TDMA scheduling and fire messages are sent
by each node in regular (periodic) intervals of $\ensuremath{T}$
seconds, followed by data packets. 

Historically, biology-inspired synchronization and desynchronization
algorithms emerged from pioneering work in \emph{pulsed-coupled oscillators}
(PCOs) \cite{mirollo1990synchronization} and integrate-and-fire models
\cite{bressloff1998desynchronization,campbell1999synchrony,kuramoto2003chemical}.
Since the original formulation of desynchronization within the context
of WSNs \cite{scaglione2010bioinspired,Degesys2007DESYNC}, several
authors extended properties of its basic reactive listening primitive
in a number of ways. Extensions towards multihop or complex network
topologies \cite{arenas2008synchronization} have been proposed via:
\emph{(i)} including neighboring information in fire messages \cite{degesys2008towards},
\emph{(ii)} low-complex graph theory methods \cite{motskin2009lightweight,kang2009localized},
and \emph{(iii)} broadcast/reception of only a limited number of beacon
messages to/from the immediate phase neighbors \cite{mutazono2011energy,de2012localized,WangTSPstatisticalPCO,muhlberger2013analyzing}.
The effects of node mobility in desynchronization were discussed in
recent work \cite{settawatcharawanit2012v}. Synchronization and desynchronization
methods with limited listening or limited beacon broadcasts were also
proposed recently for increased energy efficiency in WSN designs \cite{cornejo2010deploying,WangTSP_PCO_Energy,klinglmayr2012guaranteeing,nishimura2012probabilistic,arenas2008synchronization}.
Under the knowledge of the total number of nodes, it was shown that
maintaining one node with fixed beaconing (i.e., an ``anchored''
node) \cite{lien2012anchored} allows for faster convergence to TDMA.
Other works focused on modifications to the basic desynchronization
to allow for TDMA with: low-complexity scheduling \cite{peterhong2009pulse},
unequal slot sizes \cite{scaglione2010bioinspired,pagliari2011scalable},
as well as scheduling under discrete resources (non-continuous time)
\cite{ashkiani2012discrete}. Finally, in our recent work \cite{buranpanichkit2012distributed}
we proposed a time-frequency extension of the desynchronization process
in order to achieve increased bandwidth efficiency and allow for low-complex
distributed coordination across the multiple channels supported by
the IEEE 802.15.4 standard for WSN communications.

In all these works, the number of convergence iterations required
until the steady state plays a crucial role in latency, energy and
bandwidth efficiency of WSN deployments based on desynchronization.
For example, the required convergence iterations were a key issue
in the simulations and experiments of several desynchronization-based
systems \cite{ashkiani2012discrete,lien2012anchored,degesys2008towards,scaglione2010bioinspired,Degesys2007DESYNC}.
Beyond single-channel desychronization, convergence iterations comprise
a crucial parameter in the convergence delay of multichannel desynchronization
\cite{buranpanichkit2012distributed}, which is an important factor
in the energy consumption of practical deployments \cite{BesbesTWC2013}.
Finally, deriving estimates for the convergence iterations forms a
crucial step in the analytic understanding of how the system evolves
from random initial conditions to the desynchronized steady state
\cite{lien2012anchored}.

\subsection{Related Work}

It is well known that deriving \emph{closed-form} estimates for the
required convergence iterations to SState is hard \cite{leidenfrost2009firefly,Degesys2007DESYNC,patel2007desync,lien2012anchored}
due to the non-deterministic aspects of the desynchronization process,
namely, the random initial condition of the phase of each node and
the random perturbations in the firing order of nodes due to noise.
Therefore, existing works focus on \emph{order--of--convergence} \cite{Degesys2007DESYNC}
or \emph{lower bounds of convergence} \emph{iterations} of desynchronization,
proven or conjectured via experimentation \cite{scaglione2010bioinspired,patel2007desync,ashkiani2012discrete,lien2012anchored}.
Moreover, these works consider only the noise-free case. 

While order--of--convergence estimates provide a coarse (asymptotic)
characterization of the convergence, they do not predict the expected
number of iterations required for desynchronization to converge to
the SState. On the other hand, the existing \emph{lower bounds} on
the desynchronization convergence iterations are currently given without
a characterization on their tightness to real-world experiments or
simulations. 

Instead, following a \emph{stochastic approach }yields an analytic
understanding of the behavior of the convergence of desychronization\emph{.
}Particularly,\emph{ }it leads to analytic estimates for the desynchronization
iterations that should be a close match to experiments and simulations\emph{.
}Recent work considered probabilistic approaches to analyze properties
of synchronization \cite{leidenfrost2009firefly,WangTSP_PCO_Energy,WangTSPstatisticalPCO}.
However, no previous work proposes a framework for the analytic estimation
of the expected convergence iterations required to achieve desynchronization.
{} In addition, in all related work, the models are applicable to \emph{synchronization,}
and the required differences (i.e., different phase update, reachback
response and pre-emptive message staggering \cite{leidenfrost2009firefly}
and limiting the node connectivity \cite{WangTSPstatisticalPCO,WangTSP_PCO_Energy})
do not permit a direct mapping of their experimentally-derived convergence
estimates to desynchronization systems. 

This gap in the analytic understanding of desynchronization is in
fact explicitly recognized in the related literature \cite{scaglione2010bioinspired,ashkiani2012discrete,lien2012anchored},
where it is stated that, although desynchronization algorithms are
shown to work properly by various experiments and computer simulations,
they still lack theoretical proofs for the expected iterations until
convergence to SState.

\subsection{Contribution}

In this paper, we address this issue by embracing the non-deterministic
aspects of desynchronization and proposing \emph{stochastic }(instead
of deterministic) estimates for the convergence iterations. In order
for our estimates to have wide applicability, we focus on the two
reactive listening primitives that form the basis of desynchronization
algorithms with limited listening: \emph{(i)} the \textsc{Desync}
algorithm of Degesys \emph{et al.} \cite{Degesys2007DESYNC,patel2007desync};
\emph{(ii)} PCOs with inhibitory coupling and limited listening proposed
by Pagliari \emph{et al.} \cite{scaglione2010bioinspired}. In particular,
Degesys \emph{et al.} \cite{Degesys2007DESYNC,patel2007desync} reduce
the listening interval by considering only the temporally adjacent
firing events of each node's firing and Pagliari \emph{et al.} \cite{scaglione2010bioinspired}
limit the listening interval by introducing an appropriate PCO-dynamics
function \cite{simeOne2008distributed,mirollo1990synchronization}.
This is of particular relevance to WSNs, because reductions in the
listening interval correspond to substantial reductions in the energy
consumption of wireless sensors \cite{BesbesTWC2013,mutazono2011energy,WangTSP_PCO_Energy}. 

If the total number of nodes is known, PCOs with inhibitory coupling
and limited listening have been conjectured to converge to SState
faster than the \textsc{Desync} algorithm \cite{scaglione2010bioinspired}.
Via the proposed stochastic modeling framework, we propose analytic
estimates for the number of iterations until desynchronization is
\emph{expected to have converged} to SState within a predetermined
threshold. We validate our results based on a real WSN deployment,
as well as under a simulation environment, and demonstrate the superiority
of the proposed stochastic estimates against the existing convergence
bounds in the literature \cite{scaglione2010bioinspired,ashkiani2012discrete,lien2012anchored,Degesys2007DESYNC}.

\subsection{Paper Organization}

We first review the considered reactive listening primitives in Section
II. The proposed stochastic estimates of the convergence iterations
to SState are derived in Section III. Experimental results and comparisons
are provided in Section IV. Section V presents results when using
the proposed model within two WSN TDMA systems based on desynchronization,
while Section VI provides concluding remarks.


\section{Desynchronization Primitives}

\subsection{Notations and Symbolism}

Italicized letters indicate scalars and boldface letters indicate
vectors. For vectors $\ensuremath{{\mathbf{a}}}$ and $\ensuremath{{\mathbf{b}}}$,
the circular convolution \cite{strang1996wavelets} with period $W$
is given by ($0\leq n<W$)
\[
\ensuremath{\left(\mathbf{a}\ast\mathbf{b}\right)_{W}\left[n\right]}=\sum_{m=-\infty}^{\infty}\left(a\left[m\right]\sum_{k=-\infty}^{\infty}b\left[n-m-kW\right]\right)
\]
Random variables (RVs) are represented by Greek uppercase letters,
e.g., $\ensuremath{\Phi\sim\mathcal{N}\left({\mu_{\Phi},\sigma_{\Phi}}\right)}$
or $\ensuremath{\Delta\sim\mathcal{U}\left({\mu_{\Delta},\sigma_{\Delta}}\right)}$,
with $\ensuremath{\mathcal{N}(\cdot)}$ and $\ensuremath{\mathcal{U}(\cdot)}$
reserved to indicate the normal and uniform probability density functions
(PDFs), respectively, with mean $\ensuremath{\mu_{\Phi}}$ (and $\ensuremath{\mu_{\Delta}}$)
and standard deviation $\ensuremath{\sigma_{\Phi}}$ (and $\ensuremath{\sigma_{\Delta}}$).
The mathematical operators and key concepts used in the paper are
summarized in Table \ref{tab:Notation-table.1}. 

\begin{table}
\noindent \centering{}\protect\caption{\label{tab:Notation-table.1}Mathematical Operators and Key Concepts.}
\begin{tabular}[t]{>{\centering}p{0.2\columnwidth}>{\raggedright}p{0.7\columnwidth}}
\hline 
\noalign{\vskip\doublerulesep}
\multicolumn{1}{c}{\textbf{Symbol }} & \multicolumn{1}{c}{\textbf{Definition}}\tabularnewline[\doublerulesep]
\hline 
\centering{}$\left\Vert \mathbf{v}\right\Vert $ & vector norm-2\tabularnewline
\noalign{\vskip\doublerulesep}
\centering{}$\mathbf{v}\left[n\right]$ & the $n$th element of vector ${\mathbf{v}}$, $n\geqslant0$\tabularnewline
\noalign{\vskip\doublerulesep}
\centering{}$\ensuremath{\left(\mathbf{a}\ast\mathbf{b}\right)_{W}\left[n\right]}$ & the $n$th sample of circular convolution of period $W$\tabularnewline
\noalign{\vskip\doublerulesep}
\centering{}${\text{expr}}\,\left(\text{mod1}\right)$ & modulo-1 of expression $\ensuremath{{\text{expr}}\in\mathbb{R}}$\tabularnewline
\noalign{\vskip\doublerulesep}
$a\leftarrow{\text{expr}}$ & the value of variable $a$ is updated via expression $\ensuremath{{\text{expr}}\in\mathbb{R}}$\tabularnewline
\noalign{\vskip\doublerulesep}
\centering{}$\left\lfloor u\right\rfloor $ & the largest integer that is smaller or equal to $u$\tabularnewline
\noalign{\vskip\doublerulesep}
\centering{}$\left\lceil u\right\rceil $ & the smallest integer that is larger or equal to $u$\tabularnewline
\noalign{\vskip\doublerulesep}
\centering{}${\text{Pr}}[{\text{expr}}]$ & probability of occurrence of expression $\ensuremath{{\text{expr}}\in\mathbb{R}}$\tabularnewline[\doublerulesep]
\hline 
\noalign{\vskip\doublerulesep}
\textbf{Key Concept} & \centering{}\textbf{Explanation}\tabularnewline[\doublerulesep]
\hline 
\noalign{\vskip\doublerulesep}
\centering{}$W$ & total number of nodes in the desynchronization process\tabularnewline[\doublerulesep]
\noalign{\vskip\doublerulesep}
\centering{}$T$ & period of firing cycles (in seconds)\tabularnewline[\doublerulesep]
\noalign{\vskip\doublerulesep}
$\alpha$ & phase coupling constant of desynchronization\tabularnewline[\doublerulesep]
\noalign{\vskip\doublerulesep}
$b_{\text{thres}}$ & steady-state convergence threshold of desynchronization\tabularnewline[\doublerulesep]
\noalign{\vskip\doublerulesep}
\centering{}$\varphi_{\text{own}}^{\left(k\right)}$  & phase variable of ``$\text{own}$'' node (i.e., the node under consideration)
during its $k$th firing cycle \tabularnewline[\doublerulesep]
\noalign{\vskip\doublerulesep}
\centering{}$\varphi_{-i}^{\left(k\right)}$, $\varphi_{+i}^{\left(k\right)}$ & phase variables of the $i$th firing \emph{prior to} or \emph{after}
``$\text{own}$'' node's firing in its $k$th firing cycle\tabularnewline[\doublerulesep]
\hline 
\noalign{\vskip\doublerulesep}
\end{tabular}
\end{table}

\subsection{Introduction to Desynchronization }

Consider a WSN comprising $\ensuremath{W}$ fully-meshed nodes, i.e.,
every WSN node can receive message broadcasts from all other nodes.
Each node in the WSN is an ``oscillator'' that performs a task with
a period of $T$ seconds \cite{Degesys2007DESYNC}. In the beginning,
each node sets its internal timer to a random initial value between
$\left[0,T\right)$. Upon the completion of its cycle, each node broadcasts
a fire message and immediately resets its internal timer to zero.
In the steady (desynchronized) state, each node fires every $T$ seconds.
For each node, its \emph{firing cycle} comprises the time duration
in-between two sequential fire message transmissions of its own. For
each node, the percentage of the way through its $k$th firing cycle
is denoted as the node's \emph{own firing} \emph{phase} \cite{buranpanichkit2012distributed,Degesys2007DESYNC,degesys2008towards,leidenfrost2009firefly,pagliari2011scalable,peterhong2009pulse,scaglione2010bioinspired},
$\varphi_{\text{own}}^{\left(k\right)}\in\left[0,1\right)$. In order
to achieve convergence to steady state, beyond its own phase, each
node counts the percentage of time $t\in\left[0,T\right)$ from the
moments it receives fire messages from other nodes and updates its
own phase according to the desynchronization primitives detailed in
the next two subsections. Each node listens for fire message broadcasts
within a certain interval before (and possibly after) its own firing,
which is termed as the \emph{listening interval}. 

\begin{figure*}
\begin{centering}
\subfigure[DESYNC phase update]{ \includegraphics[scale=0.19]{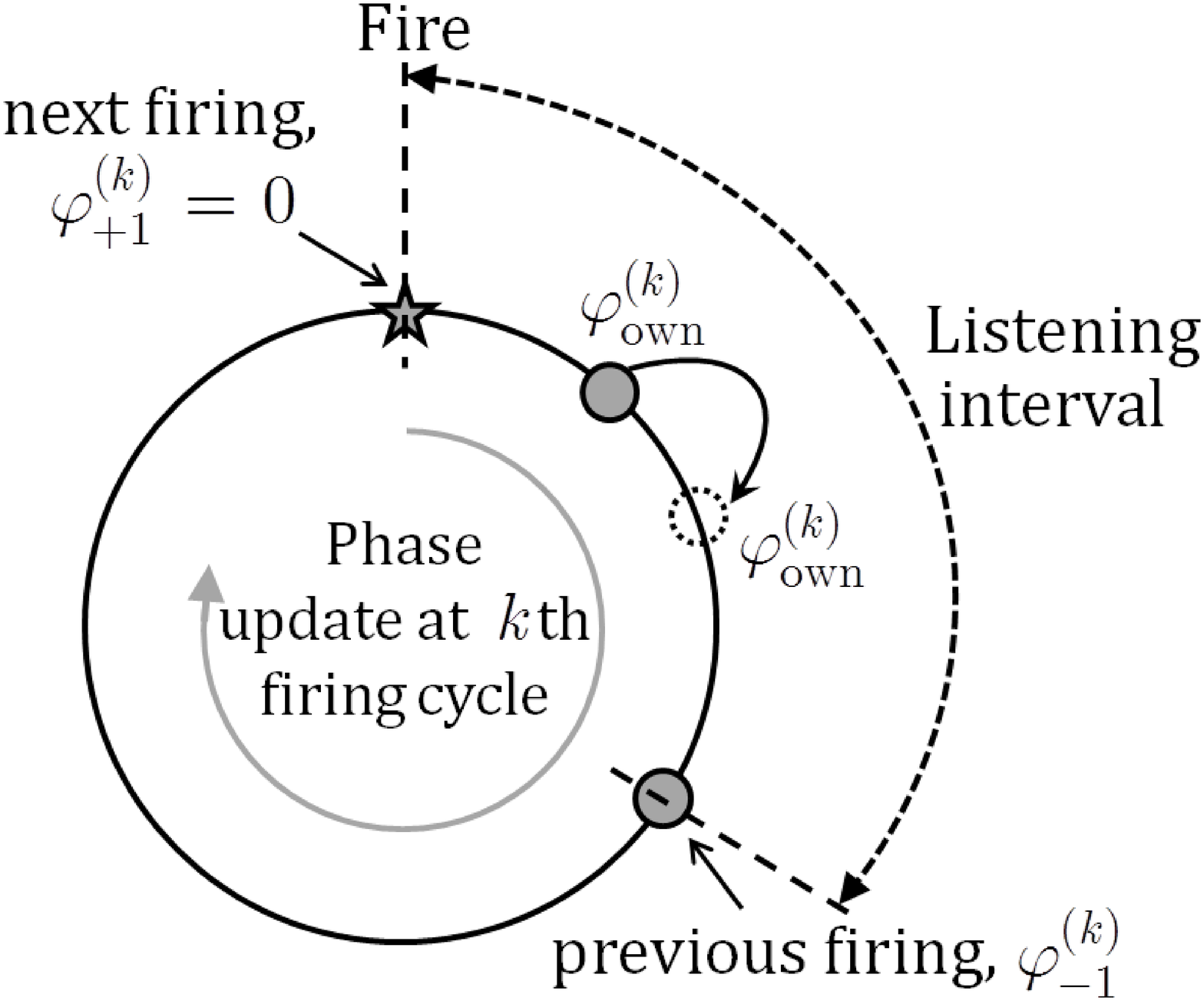}
} \subfigure[PCO-based phase update]{ \includegraphics[scale=0.19]{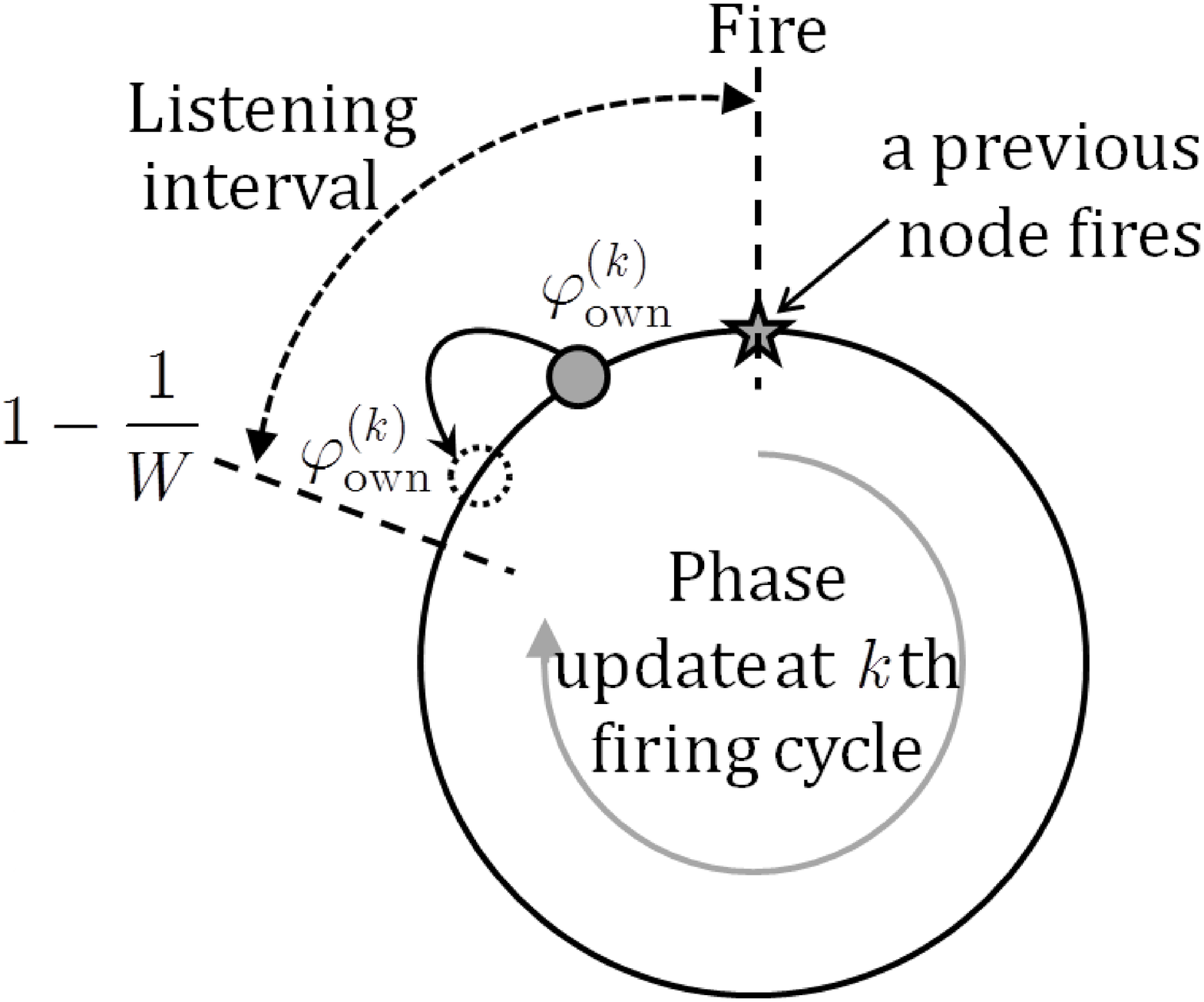}
}
\par\end{centering}

\protect\caption{The phase update of ``$\text{own}$'' node (node under consideration)
during its $k$th firing cycle when: (a) the next fire message is
received in \textsc{Desync}, with the node's listening interval defined
as the time between the fire message preceding and following its own
firing; (b) a fire message is received in PCO-based desynchronization
while the node's own phase is within its listening interval (i.e.,
when $1-\frac{1}{W}<\varphi_{\text{own}}^{(k)}<1$).}

\label{fig:Phase update} 
\end{figure*}

Following the schema of Degesys \emph{et al.} \cite{Degesys2007DESYNC},
the fire messages' phase values can be imagined as beads moving clockwise
on a ring with period $\ensuremath{T=1}$s (Fig. \ref{fig:Phase update}).
When a node's own firing phase reaches unity, i.e., top of Fig. \ref{fig:Phase update},
the node broadcasts a fire message and the node's own firing phase
is reset to zero. The figure illustrates a node's own firing phase
during its $k$th firing cycle and Fig. \ref{fig:Phase update}(a)
shows the phase of the received fire messages from the two nodes that
fired immediately before and after it, denoted by $\varphi_{-1}^{(k)}$
and $\varphi_{+1}^{(k)}$, respectively%
\footnote{In this section, we ignore the phase measurement noise and assume
each fire time can be determined precisely by all receiving nodes.
This noise is however taken into account in the modeling framework. %
}. Thus, superscripts always indicate the \textit{firing}\textit{\emph{
}}\emph{cycle} of ``$\text{own}$'' node in the WSN (i.e., of the
node under consideration) and subscripts indicate the order relative
to ``$\text{own}$'' node. The nodes corresponding to the previous
and next firings of ``$\text{own}$'' node are called \emph{phase
neighbors} and the listening interval corresponding to the $k$th
firing of ``$\text{own}$'' node is illustrated in Fig. \ref{fig:Phase update}.
For all desynchronization algorithms under consideration \cite{scaglione2010bioinspired,Degesys2007DESYNC,patel2007desync,lien2012anchored,degesys2008towards,buranpanichkit2012distributed,de2012localized,settawatcharawanit2012v}: 
\begin{enumerate}
\item the notion of phase neighbors indicates temporal adjacency of fire
messages and is independent of the nodes' physical location;
\item it is immaterial which physical sensor node is linked to which fire
message, as desynchronization is solely dependent on the received
fire message phase.
\end{enumerate}
Therefore, the analysis of this paper is presented \emph{from the
viewpoint of any single node in the WSN} \cite{kang2009localized,muhlberger2013analyzing,degesys2008towards}
and does not need to discern the specific firing order of all nodes
in the network, which in fact may not be constant during the desynchronization
convergence process.

\subsection{\textsc{Desync} }

In this approach, each node updates its firing phase \textit{once
}within each of its firing cycles, at the moment when the next fire
message is received. As shown in Fig. \ref{fig:Phase update}(a),
each node listens for the fire message preceding and following its
own fire message broadcast. Hence, the duration of each node's listening
interval depends on the relative times inbetween these two messages.
It then performs its phase update during its $k$th firing cycle based
on the phases of these two received messages, i.e., $\varphi_{-1}^{\left(k\right)}$
and $\varphi_{+1}^{\left(k\right)}$. Specifically, each node's phase
update moves its own firing phase towards the middle of the listening
interval%
\footnote{Since (\ref{eq:DESYNC phase update}) is applied at the moment when
the next firing is received, we have: $\varphi_{+1}^{(k)}=0$, as
seen in Fig. \ref{fig:Phase update}(a); however, we include $\varphi_{+1}^{(k)}$
in (\ref{eq:DESYNC phase update}) to clarify that the operation of
\textsc{Desync} depends on both the previous and next firing phase. %
}. The phase update of \textsc{Desync} during the $k$th firing cycle
is expressed by \cite{Degesys2007DESYNC,patel2007desync}
\begin{equation}
\varphi_{\text{own}}^{(k)}\leftarrow\left[\left(1-\alpha\right)\varphi_{\text{own}}^{(k)}+\frac{\alpha}{2}\left(\varphi_{-1}^{(k)}+\varphi_{+1}^{(k)}\right)\;\;\left(\text{mod}1\right)\right],\label{eq:DESYNC phase update}
\end{equation}
where $\alpha\in(0,1)$ denotes the \emph{phase-coupling constant}
that controls the speed of the phase adaptation. Previous work \cite{Degesys2007DESYNC,patel2007desync}
showed that the reactive listening primitive of (\ref{eq:DESYNC phase update})
disperses all fire message broadcasts at intervals of $\frac{T}{W}$.
Thus, after $k_{{\text{SState}}}$ firing cycles, the \textsc{Desync}
algorithm leads to fair TDMA scheduling, where all fire messages in
the network are periodic and the phase update of (\ref{eq:DESYNC phase update})
leads to convergence to SState, expressed by
\begin{align}
\exists k_{\text{SState}}\in\mathbb{N}^{\star}\;\text{s.t.} & \;\forall k\geq k_{\text{SState}}:\;\left|\varphi_{\text{own}}^{(k)}-\varphi_{-1}^{(k)}-\frac{1}{W}\right|\leq b_{{\text{thres}}},\label{eq:check convergence}
\end{align}
with $b_{{\text{thres}}}$ the preset convergence threshold, typically
\cite{buranpanichkit2012distributed,Degesys2007DESYNC,degesys2008towards,pagliari2011scalable,peterhong2009pulse,scaglione2010bioinspired}
$b_{{\text{thres}}}\in[0.001,0.020]$. In a practical WSN deployment
\cite{BesbesTWC2013,buranpanichkit2012distributed,muhlberger2013analyzing,WangTSPstatisticalPCO,Degesys2007DESYNC},
each node would check that the condition of \eqref{eq:check convergence}
holds for several consecutive firing cycles beyond $k=k_{\text{SState}}$
(e.g., five cycles) and would then declare that convergence has been
achieved. 

In steady state, each node transmits data packets for $T\left(\frac{1}{W}-b_{{\text{thres}}}\right)$
seconds immediately following its fire message broadcast (which limits
the maximum number of nodes supported under TDMA to less than $\left\lfloor \frac{1}{b_{{\text{thres}}}}\right\rfloor $).
The only overhead stems from the fire message broadcasts, which are
very short packets (just two bytes in our implementation). 

Assuming negligible propagation delay and error-free detection of
messages, it has been conjectured via simulations \cite{scaglione2010bioinspired,Degesys2007DESYNC}
that convergence requires iterations of the order
\begin{equation}
k_{\text{DESYNC,[3][4]}}\sim O\left(\frac{1}{\alpha}W^{2}\ln\frac{1}{b_{\text{thres}}}\right).\label{eq:conjecture of DESYNC TDMA}
\end{equation}
Even though the order estimate \eqref{eq:conjecture of DESYNC TDMA}
gives a coarse characterization for the convergence iterations, it
cannot provide the expected number of iterations until convergence
to SState is achieved. Moreover, real-world experimentation with TelosB
and iMote2 motes \cite{Degesys2007DESYNC,buranpanichkit2012distributed}
under fixed $\alpha$ and $b_{\text{thres}}$ values show that the
measured number of iterations until convergence to SState is not proportional
to $W^{2}$.

\subsection{PCO-based Inhibitory Coupling }

PCO-based desynchronization with inhibitory coupling updates each
node's own firing phase according to the received fire messages within
a certain interval of its firing cycle \cite{scaglione2010bioinspired}.
This is termed as the \emph{listening interval} and its positioning
within the $k$th firing cycle of a node is illustrated Fig. \ref{fig:Phase update}(b).
As such: \textit{(i)} the phase of the node's own firing changes after
a fire message from a previous phase-neighbor is received within the
listening interval (i.e., unlike \textsc{Desync}, a varying number
of phase updates may occur within \emph{each} firing cycle); \textit{(ii)}
knowledge of the total number of nodes ($W$) is required \cite{scaglione2010bioinspired}.
Hence, a phase update during the $k$th firing cycle of PCO-based
desynchronization of a node \cite{scaglione2010bioinspired} is performed
at $\varphi_{\text{own}}^{(k)}\times T$ seconds after the node's
last fire message broadcast, $1-\frac{1}{W}<\varphi_{\text{own}}^{(k)}<1$:
\begin{equation}
\varphi_{\text{own}}^{(k)}\leftarrow\left[\left(1-\alpha\right)\varphi_{\text{own}}^{(k)}+\alpha\left(1-\frac{1}{W}\right)\;\;\left(\text{mod}1\right)\right],\label{eq:PCO phase update}
\end{equation}
where $\alpha\in(0,1)$ the phase-coupling constant controlling the
speed of the phase adaptation. All fire messages received outside
the listening interval $\left(1-\frac{1}{W},1\right)$ are simply
ignored. After $k_{{\text{SState}}}$ firing cycles, (\ref{eq:PCO phase update})
has been shown to converge to dispersed fire message broadcasts received
at intervals of $\frac{1}{W}\times T$ seconds \cite{scaglione2010bioinspired},
i.e., (\ref{eq:check convergence}) holds under convergence to SState.
Hence, like in the \textsc{Desync} case, once fair TDMA is achieved,
the only overhead stems from the short fire message broadcasts. 

Assuming negligible propagation delay, error-free detection of broadcast
messages and that $1-\frac{1}{W}>\alpha$, it has been shown \cite{scaglione2010bioinspired}
that the number of firing cycles for convergence is lower bounded
by
\begin{equation}
k_{\text{PCO,[3]}}\geq\left\lceil \frac{\ln b_{\text{thres}}-\ln\left[2+\frac{2}{\alpha^{W}\left(1-\alpha\right)}\right]}{\ln\left(1-\alpha\right)+\ln W}\right\rceil .\label{eq:conjecture of PCO}
\end{equation}
Nevertheless, the tightness of \eqref{eq:conjecture of PCO} has neither
been proven nor demonstrated via real-world experiments or simulation
results.

\section{Proposed Stochastic Modeling of \textsc{Desync} and PCO-based Desynchronization
with Inhibitory Coupling}

\textbf{Assumption 1}\emph{ }\textbf{(Fully-meshed Topology):} \emph{We
consider fully-meshed networks where each node can directly receive
the fire message broadcasts from all other nodes. }

When performing the first phase update of \eqref{eq:DESYNC phase update}
and \eqref{eq:PCO phase update}, each node's own firing phase, as
well as the phases of all received fire messages from its phase neighbors,
are modeled by independent random variables that are uniformly distributed
in $\left[0,1\right)$. This is formally stated in the following assumption.

\textbf{Assumption 2 (Phase Model):}\emph{ For every node under consideration
(``}$\text{own}$\emph{'' node) and its phase neighbors, their firing
phase during the first phase update is modeled by:} $\forall i\in\left\{ -1,\text{own},+1\right\} :\;\Phi_{i}^{(1)}\sim{\text{P}}_{\Phi_{i}^{(1)}}$\emph{,
with}%
\footnote{The use of the modulo operator in \eqref{eq:Definition_2} is imposed
because, by definition, we must ensure $\varphi_{i}\in\left[0,1\right)$.%
}
\begin{equation}
{\text{P}}_{\Phi_{i}^{(1)}}=\mathcal{U}\left(\mu_{\Phi_{i}^{(1)}},\sigma_{\Phi^{(0)}}\right)\;\;\left(\text{mod1}\right).\label{eq:Definition_2}
\end{equation}
\emph{We define the mean times of successive phase updates to be equidistant,
which, for the }\textsc{Desync}\emph{ update of (\ref{eq:DESYNC phase update}),
is expressed as }
\begin{equation}
\mu_{\Phi_{-1}^{(1)}}-\mu_{\Phi_{\text{own}}^{(1)}}=\frac{1}{W},\;\;\;\mu_{\Phi_{\text{own}}^{(1)}}-\mu_{\Phi_{+1}^{(1)}}=\frac{1}{W}\label{eq:Definition_2_part_1}
\end{equation}
\emph{and for the PCO update of (\ref{eq:PCO phase update}) is stated
by}
\begin{equation}
\mu_{\Phi_{\text{own}}^{(1)}}=1-\frac{1}{W}.\label{eq:PCO update}
\end{equation}
\emph{In the beginning of the desynchronization process, all fire
message broadcasts are completely uncoordinated (random), i.e., $\sigma_{\Phi^{(1)}}=\frac{1}{\sqrt{12}}$.} 

We remark that there is no loss of generality from the assumption
of equidistant means of (\ref{eq:Definition_2_part_1}) and (\ref{eq:PCO update})
as the modulo operator of (\ref{eq:Definition_2}) ensures that the
PDFs wrap around one such that for every node and $\forall i\in\left\{ -1,\text{own},+1\right\} :\;\Phi_{i}^{(1)}$
are always uniformly-distributed between $\left[0,1\right)$ irrespective
of the assumed mean values. However, we opt for the use of (\ref{eq:Definition_2_part_1})
and (\ref{eq:PCO update}) as this facilitates the mathematical exposition
of the proposed estimates.

Our estimates of the convergence iterations for \textsc{Desync} and
PCO-based desynchronization assume that each phase in (\ref{eq:DESYNC phase update})
and (\ref{eq:PCO phase update}) is contaminated by white noise due
to the varying propagation, mobility and node processing delays of
a WSN environment. This is captured in the following assumption. 

\textbf{Assumption 3}\textbf{\emph{ }}\textbf{(Measurement Noise Model)}\textbf{\emph{:}}
\emph{All phase values in the update of (\ref{eq:DESYNC phase update})
or (\ref{eq:PCO phase update}) are contaminated by additive noise,
modeled by an independent, zero-mean, uniformly-distributed, random
variable, $\Delta\sim\mathcal{U}\left(0,\sigma_{\Delta}\right)$.}

The assumption of uniform distribution for the measurement noise,
as well as the value of the standard deviation $\sigma_{\Delta}$
will be derived experimentally, as they incorporate the effects of
interference, wireless propagation and processing delays that can
only be inferred via measurements from a real setup.

Due to the measurement noise and the interaction between fire message
broadcasts, for each phase update of each node's firing phase, $\varphi_{\text{own}}^{\left(k\right)}$,
during its $k$th firing cycle, the PDF of $\Phi_{\text{own}}^{(k)}$,
${\text{P}}_{\Phi_{\text{own}}^{(k)}}$, changes after applying (\ref{eq:DESYNC phase update})
or (\ref{eq:PCO phase update}). Consequently, this changes the probability
of convergence to SState, as follows:
\begin{eqnarray}
\Pr\left[{\left|{\Phi_{\text{own}}^{(k)}-\mu_{\Phi_{\text{own}}^{(k)}}}\right|\leq b_{{\text{thres}}}}\right] \nonumber
\\
 = & \int_{-b_{{\text{thres}}}}^{b_{{\text{thres}}}}{\text{P}}_{\Phi_{\text{own}}^{(k)}}\left({u-\mu_{\Phi_{\text{own}}^{(k)}}}\right)du\nonumber \\
 = & \ensuremath{{\text{erf}}\left({\frac{{b_{{\text{thres}}}}}{{\sqrt{2}\sigma_{\Phi_{\text{own}}^{(k)}}}}}\right)},\label{eq:probability of convergence to SS}
\end{eqnarray}
where ${\text{erf}}\left(u\right)=\frac{2}{\sqrt{\pi}}\int_{0}^{u}e^{-t^{2}}dt$
is the error function \cite{papoulis1989probability}. Notice that
(\ref{eq:probability of convergence to SS}) holds under the assumption
that ${\text{P}}_{\Phi_{\text{own}}^{(k)}}$ converges to a normal
distribution for both \textsc{Desync} and PCO-based desynchronization,
which, as the next two subsections will show, turns out to be the
case. We therefore use a stochastic criterion for convergence based
on the confidence intervals of the normal distribution \cite{papoulis1989probability}.
By defining the confidence coefficient
\begin{equation}
c_{{\text{conf}}}=\ensuremath{\Pr\left[{\left|{\Phi_{\text{own}}^{(k)}-\mu_{\Phi_{\text{own}}^{(k)}}}\right|\leq b_{{\text{thres}}}}\right]},\;\;0<c_{{\text{conf}}}<1,
\end{equation}
and replacing in (\ref{eq:probability of convergence to SS}), we
get
\begin{equation}
\sigma_{\Phi_{\text{own}}^{(k)}}=\frac{b_{{\text{thres}}}}{\sqrt{\text{2}}\times{\text{erf}}^{-1}\left({c_{{\text{conf}}}}\right)},\label{eq:confidence coefficient}
\end{equation}
 with ${\text{erf}}^{-1}(u)$ the inverse error function that can
be computed by its Maclaurin series
\begin{equation}
{\text{erf}}^{-1}(u)=\frac{1}{2}\sqrt{\pi}\left({u+\frac{\pi}{{12}}u^{3}+\frac{{7\pi^{2}}}{{480}}u^{5}+\ldots}\right).
\end{equation}

Thus, (\ref{eq:confidence coefficient}) becomes the mechanism for
defining the phase update iteration leading to SState. Specifically,
we determine the firing cycle $k_{{\text{SState}}}$ for which $\sigma_{\Phi_{\text{own}}^{(k_{{\text{SState}}})}}$
is closest to the right-hand side of \eqref{eq:confidence coefficient}.
That is, we determine the firing cycle leading to convergence to SState
with probability that closely matches $c_{{\text{conf}}}$, which
is our (predetermined) confidence.

\textbf{Definition 1}\textbf{\emph{ }}\textbf{(Steady State):} \emph{We
define a desynchronization primitive as being in steady state with
}$c_{{\text{conf}}}\times100$\emph{\% confidence, at the} $k_{{\text{SState}}}$\emph{th
firing cycle, }$0<c_{{\text{conf}}}<1$,\emph{ where}
\begin{equation}
k_{{\text{SState}}}=\arg\mathop{\min}\limits _{\forall k\in\mathbb{N}}\left|{\sigma_{\Phi_{\text{own}}^{(k)}}-\frac{b_{{\text{thres}}}}{\sqrt{2}\times{\text{erf}}^{-1}\left({c_{{\text{conf}}}}\right)}}\right|,
\end{equation}
\emph{with }$\sigma_{\Phi_{\text{own}}^{(k)}}$ \emph{the standard
deviation of a node's own firing phase PDF at the application of the
updates of (\ref{eq:DESYNC phase update}) or (\ref{eq:PCO phase update})
during its $k$th firing cycle.}

Since $\sigma_{\Phi_{\text{own}}^{\left(k\right)}}$ is affected by
measurement noise, in order for the system to remain in the converged
state indefinitely, the threshold for the convergence, $b_{{\text{thres}}}$,
must be set according to the (estimated) $\sigma_{\Delta}$. Conversely,
we can treat the entire desynchronization process as a ``black box\textquotedblright{}
system and estimate $\sigma_{\Delta}$ by measuring the phase deviation
from the mean obtained when performing the update of (\ref{eq:DESYNC phase update})
or (\ref{eq:PCO phase update}) during SState. This will be demonstrated
in the experimental section.

\subsection{Modeling of \textsc{Desync} Convergence }

\textbf{Proposition 1.} \emph{Under Assumptions 1--3, the expected
number of firing cycles for the }\textsc{Desync}\emph{ phase update
of (\ref{eq:DESYNC phase update}) to converge according to Definition
1 is}
\begin{equation}
k_{{\text{desync}}}=\arg\mathop{\min}\limits _{\forall k\in\mathbb{N}}\left|{\sigma_{\text{desync},k}-\frac{b_{{\text{thres}}}}{\sqrt{2}\times{\text{erf}}^{-1}\left({c_{{\text{conf}}}}\right)}}\right|,\label{eq:number of firing cycles for DESYNC}
\end{equation}
 \emph{with}
\begin{equation}
\sigma_{{\text{desync}},k}=\sqrt{\left\Vert {{\mathbf{v}}_{W}^{(k)}}\right\Vert ^{2}\sigma_{\Phi^{(1)}}^{2}+\sum\limits _{j=1}^{k}{\left\Vert {{\mathbf{v}}_{W}^{(j)}}\right\Vert ^{2}\sigma_{\Delta}^{2}}},\label{eq:std of DESYNC}
\end{equation}
\begin{equation}
{\mathbf{v}}=\begin{bmatrix}\frac{\alpha}{2} & 1-\alpha & \frac{\alpha}{2}\end{bmatrix},\label{eq:zero padding V}
\end{equation}
\emph{and}
\begin{center}
${\mathbf{v}}_{W}^{(j)}=\underset{j\text{ times}}{\underbrace{\left({\mathbf{v}}\ast\ldots\ast{\mathbf{v}}\right)_{W}}},$
\par\end{center}
\emph{being the vector produced by $j$ consecutive circular convolutions
of period $W$.} 
\begin{IEEEproof}
Consider any single node in the WSN. We denote the initial phase random
variables corresponding to the node under consideration by the $1\times W$
vector
\begin{equation}
\mathbf{{\color{black}\Phi}}^{\left(1\right)}=\begin{bmatrix}\ldots & \Phi_{-1}^{\left(1\right)} & \Phi_{\text{own}}^{\left(1\right)} & \Phi_{+1}^{\left(1\right)} & \ldots\end{bmatrix}.\label{eq:phi-vector}
\end{equation}
The corresponding additive measurement noise sources {[}independent
identically distributed (i.i.d.) random variables from Assumption
3{]} are denoted by the $1\times W$ vector
\begin{equation}
\mathbf{\Delta}^{\left(1\right)}=\begin{bmatrix}\ldots & \Delta_{-1}^{\left(1\right)} & \Delta_{\text{own}}^{\left(1\right)} & \Delta_{+1}^{\left(1\right)} & \ldots\end{bmatrix}.\label{eq:delta-vector}
\end{equation}
Evidently, the number of elements before and after $\Phi_{\text{own}}^{\left(1\right)}$
and $\Delta_{\text{own}}^{\left(1\right)}$ in (\ref{eq:phi-vector})
and (\ref{eq:delta-vector}) depends on how many fire message broadcasts
(firings) precede or follow the node's own firing during its initial
firing cycle. The first phase update of (\ref{eq:DESYNC phase update})
is expressed stochastically as
\begin{eqnarray}
\Phi_{\text{own}}^{\left(1\right)} & \leftarrow & \left[\left(1-\alpha\right)\left(\Phi_{\text{own}}^{\left(1\right)}+\Delta_{\text{own}}^{\left(1\right)}\right)\right.\label{eq:Desync_prob_phase_update}\\
 & + & \left.\frac{\alpha}{2}\left(\Phi_{-1}^{\left(1\right)}+\Delta_{-1}^{\left(1\right)}+\Phi_{+1}^{\left(1\right)}+\Delta_{+1}^{\left(1\right)}\right)\;\;\left(\text{mod}1\right)\right].\nonumber \end{eqnarray}
Notice that \eqref{eq:Desync_prob_phase_update} imposes that the
statistics of $\Phi_{-1}^{(1)}$ and $\Phi_{+1}^{(1)}$ correspond
to the initial firing cycle (Assumption 2). This is because, during
each phase update, we do not take into account nodes' phase updates
that may have been carried out \emph{during} the first firing cycle.
This corresponds to the operational form of the \textsc{Desync} algorithm
(i.e., ``\textsc{Desync} stale'' of \cite{Degesys2007DESYNC,degesys2008towards}).
It is straightforward to derive from \eqref{eq:Desync_prob_phase_update}
that $\mu_{\Phi_{\text{own}}^{\left(1\right)}}$ remains unchanged
after the first phase update, while the standard deviation is modified
to:
\begin{equation}
\sigma_{\text{desync},1}=\left\Vert {\mathbf{v}}\right\Vert \sqrt{(\sigma_{\Phi^{\left(1\right)}}^{2}+\sigma_{\Delta^{\left(1\right)}}^{2})}.
\end{equation}
Furthermore, by writing \eqref{eq:Desync_prob_phase_update} for all
the phase elements of $\mathbf{{\color{black}\Phi}}^{\left(1\right)}$
given in (\ref{eq:phi-vector}), we reach 
\begin{equation}
\mathbf{\Phi}^{\left(1\right)}\leftarrow\left[\left[\mathbf{v\ast}\left(\mathbf{\Phi}^{\left(1\right)}+\mathbf{\Delta}^{\left(1\right)}\right)\right]_{W}\;\;\left(\text{mod}1\right)\right].\label{eq:first iteration of DESYNC phase update}
\end{equation}
 The circular convolution performs periodic extension of the phase
and noise vectors of (\ref{eq:phi-vector}) and (\ref{eq:delta-vector}),
which corresponds to the circular dependency between consecutive firing
cycles%
\footnote{For the special cases of $W\in\{2,3,4\}$ nodes, we set $W=5$ in
the circular convolution of (\ref{eq:zero padding V}) to avoid erroneous
overlapping within ${\mathbf{v}}^{(j)}$ due to the short period of
the circular convolution.%
}.

Generalizing (\ref{eq:first iteration of DESYNC phase update}) to
the $k$th firing cycle, we reach
\begin{eqnarray}
\mathbf{\Phi}^{\left(k\right)} & = & \left(\mathbf{v}_{W}^{\left(k\right)}\ast\mathbf{\Phi}^{\left(1\right)}\right)_{W}\label{eq:k-th iteration of DESYNC update}\\
 & + & \sum\limits _{j=1}^{k}\left(\mathbf{v}_{W}^{\left(j\right)}\ast\mathbf{\Delta}^{\left(j\right)}\right)_{W}\;\left(\text{mod}1\right),\nonumber \end{eqnarray}
where $\mathbf{\Delta}^{\left(j\right)}$ is the i.i.d. measurement
noise vector per iteration. Therefore, we obtain: $\mu_{\Phi_{\text{own}}^{\left(k\right)}}=\mu_{\Phi_{\text{own}}^{\left(1\right)}}$
and $\sigma_{{\text{desync}},k},$ shown in (\ref{eq:std of DESYNC}).
It can now be shown that $\Phi_{\text{own}}^{\left(k\right)}$ becomes
normally distributed after a few firing cycles (see Appendix \ref{sec:Appendix-I}),
i.e., 
\begin{equation}
\Phi_{\text{own}}^{\left(k\right)}\sim\mathcal{N}\left(\mu_{\Phi_{\text{own}}^{\left(1\right)}},\sigma_{{\text{desync}},k}\right)\;\;\left(\text{mod}1\right).
\end{equation}
Hence, we reach (\ref{eq:number of firing cycles for DESYNC}) for
convergence under Definition 1. 
\end{IEEEproof}
Proposition 1 shows how $k_{{\text{desync}}}$ is affected by $\alpha$
as well as by the initial conditions and the noise assumptions expressed
by $\sigma_{\Phi^{\left(1\right)}}$ and $\sigma_{\Delta}$ in Assumptions
2 and 3, respectively. Interestingly, according to \eqref{eq:number of firing cycles for DESYNC},
the number of nodes, $W$, does not appear to influence the convergence
to the steady state. This is in contrast to the conjecture of Degesys
\emph{et al}. \cite{Degesys2007DESYNC,scaglione2010bioinspired} given
by \eqref{eq:conjecture of DESYNC TDMA}. Nevertheless, experimental
results given in the next section will demonstrate that real-world
WSNs, as well as simulation results, are in agreement with Proposition
1.

\subsection{Modeling of PCO-based Convergence}

\textbf{Proposition 2.} \emph{Under Assumptions 1--3, the expected
number of firing cycles for the PCO-based phase update of (\ref{eq:PCO phase update})
to converge according to Definition 1 is}
\begin{eqnarray}
k_{{\text{PCO}}} & = & \arg\mathop{\min}\limits _{\forall k\geq2}\left|\sum\limits _{l=2}^{k}\left[\text{erf}\left(\frac{\left\lfloor \frac{W}{2}\right\rfloor +1}{W\sigma_{\text{PCO},l}\sqrt{2}}\right)\right.\right.\label{eq:firing cycles for PCO convergence}\\
 & - & \left.\left.\frac{1}{2}\text{erf}\left(\frac{1}{W\sigma_{\text{PCO},l}\sqrt{2}}\right)\right]+1-\frac{1}{W}-l_{{\text{SSupd}}}\right|,\nonumber \end{eqnarray}
 \emph{with} 
\begin{equation}
l_{{\text{SSupd}}}{\text{ = }}\arg\mathop{\min}\limits _{\forall l\in\mathbb{N}}\left|{\sigma_{{\text{PCO}},l}-\frac{b_{{\text{thres}}}}{\sqrt{2}\times{\text{erf}}^{-1}\left({c_{{\text{conf}}}}\right)}}\right|,\label{eq:number of phase update of PCO}
\end{equation}
 \emph{and} $\forall l\in\mathbb{N}$,
\begin{equation}
\sigma_{{\text{PCO}},l}{\text{ = }}\sqrt{(1-\alpha)^{2l}\sigma_{\Phi^{\left(1\right)}}^{2}+\frac{{(\alpha-1)^{2}}}{{\alpha(\alpha-2)}}\left[{(1-\alpha)^{2l}-1}\right]\sigma_{\Delta}^{2}}.\label{eq:std of PCO}
\end{equation}
\begin{IEEEproof}
We separate the proof into three parts, based on the temporal evolution
of the convergence process. We present here the main part of the proof,
which comprises the analysis of the first firing cycle and the derivation
of the $l$th phase update of a node during its $k$th firing cycle
in PCO, while the remaining details to complete the proof are given
in Appendix \ref{sec:Appendix-II}. 

\textbf{First firing cycle:} Consider any single node in the WSN.
The expected number of phase updates it will perform within its first
firing cycle is equal to the number of fire message broadcasts (firings)
expected to be heard by the node within $\varphi_{\text{own}}^{\left(1\right)}\in(1-\frac{1}{W},1)$,
which is 
\begin{equation}
\sum\limits _{j=1}^{W-1}{j\left({\begin{array}{c}
{W-1}\\
j
\end{array}}\right)\left({\frac{1}{W}}\right)^{j}\left({1-\frac{1}{W}}\right)^{W-1-j}}=1-\frac{1}{W}.
\end{equation}
This stems from the binomial theorem, since Assumption 2 mandates
that the initial phase of each node is uniformly distributed within
$\left[0,1\right)$. Via (\ref{eq:PCO phase update}), a phase update
during the first firing cycle of a node can be expressed stochastically
as
\begin{equation}
\Phi_{\text{own}}^{\left(1\right)}\leftarrow\left[\left(1-\alpha\right)\left(\Phi_{\text{own}}^{\left(1\right)}+\Delta_{\text{own}}^{(1)}\right)+\alpha\left(1-\frac{1}{W}\right)\;\left(\text{mod1}\right)\right],\label{eq:first phase update of PCO}
\end{equation}
with $\Delta_{\text{own}}^{\left(1\right)}$ the random variable modeling
the measurement noise (Assumption 3) of the node's own phase. From
(\ref{eq:PCO update}) and (\ref{eq:first phase update of PCO}) we
obtain 
\begin{equation}
\mu_{\Phi_{\text{own}}^{\left(1\right)}}=1-\frac{1}{W},\label{eq:PCO equidistant mean values}
\end{equation}
i.e., the mean values of successive fire message updates remain equidistant
after the first firing cycle of a node. The standard deviation of
$\Phi_{\text{own}}^{\left(1\right)}$ after the update of \eqref{eq:first phase update of PCO}
is
\begin{equation}
\sigma_{{\text{PCO}},1}=\left({1-\alpha}\right)\sqrt{\sigma_{\Phi^{\left(1\right)}}^{2}+\sigma_{\Delta}^{2}}.\label{eq:std of first iteration of PCO phase update}
\end{equation}
Generalizing \eqref{eq:first phase update of PCO} to the $l$th phase
update of the $k$th firing cycle of a node, leads to 
\begin{eqnarray}
\Phi_{\text{own}}^{\left(k\right)} & = & \left(1-\alpha\right)^{l}\Phi_{\text{own}}^{\left(1\right)}+\sum_{j=1}^{l}\left(1-\alpha\right)^{j}\Delta_{\text{own}}^{(l-j+1)}\label{eq:PCO phase RV of kth phase update}\\
 & + & \alpha\left(1-\frac{1}{W}\right)\sum_{j=0}^{l-1}\left(1-\alpha\right)^{j}\;\;(\text{\text{mod1),}}\nonumber \end{eqnarray}
with $\Delta_{\text{\text{own}}}^{(l-j+1)}$ i.i.d. random variables,
each stemming from Assumption 3. The mean of $\Phi_{\text{own}}^{(k)}$
is given by (\ref{eq:PCO equidistant mean values}) and its standard
deviation is given by (\ref{eq:std of PCO}). Similarly to Proposition
1, it can be shown that $\Phi_{\text{own}}^{\left(k\right)}$ becomes
a normally-distributed random variable after a few phase updates (see
Appendix \ref{sec:Appendix-I}). We can thus reach convergence under
Definition 1 for $l$ given by (\ref{eq:number of phase update of PCO}).
However, given that in PCO-based desynchronization the number of phase
updates per firing cycle is \emph{not} fixed, i.e., in general, $l\neq k$,
in order to derive the \emph{expected number of firing cycles} until
a node converges to steady state, we need to derive the expected number
of phase updates after each firing cycle. We can then match the number
of phase updates expected to take place until convergence to the corresponding
number of firing cycles. The details of this process and the proof
of \eqref{eq:firing cycles for PCO convergence} are given in Appendix
\ref{sec:Appendix-II}.
\end{IEEEproof}
Proposition 2 shows that $k_{{\text{PCO}}}$ is affected by $\alpha$,
as well as by the initial conditions and the noise assumptions, expressed
by $\sigma_{\Phi^{(1)}}$ and $\sigma_{\Delta}$ in Assumptions 2
and 3. The total number of nodes, $W$, is also influencing the number
of iterations for convergence to the steady state. However, as it
will be shown by the next section (experiments), this effect is negligible
in practice. This is in contrast to the lower bound derived by Pagliari
\emph{et al. }\cite{scaglione2010bioinspired}, given by \eqref{eq:conjecture of PCO}.
However, the experimental results of the next section will demonstrate
that real-world WSNs, as well as simulation results, are in agreement
with Proposition 2.

\section{Experimental Validation}

For our experiments, we used iMote2 Crossbow sensors with TinyOS1.x.
All nodes use the IEEE 802.15.4 standard with the default 2.4GHz Chipcon
CC2420 wireless transceiver. We followed the TinyOS standard message
format but reduced it to two data bytes when sending fire messages.
Similar to prior work \cite{Degesys2007DESYNC}, we set the backoff
time to 1.2ms.

\subsection{Standard Deviation of the Phase Measurement Noise}

The test environment was a standard university laboratory room. Our
approach for measuring $\sigma_{\Delta}$ was carried out as follows:\textit{
(i) }we implemented the \textsc{Desync} and PCO-based desynchronization
in TinyOS nesC code as described in Section II;\textit{ (ii)} we set
$\alpha=0.95$ to ensure maximum coupling strength and $T=1$s (this
period value was used in all our experiments) and \textit{(iii)} we
measured the oscillatory behavior of each node's phase after the
WSN was left operating for a prolonged interval of time (during which
nodes were occasionally moved within the test area) to ensure convergence
to SState. The statistics of the oscillatory phase behavior, observed
via this experiment, express the cummulative effects of interference,
mobility and clock drift amongst nodes. This approach is easy to replicate
under any real-world WSN setup involving varying levels of interference
or node mobility \cite{nakano2011biologically,lien2012anchored,settawatcharawanit2012v}.

For both algorithms, we found the standard deviation of the oscillating
phase amplitude around the SState value of each node's phase to be
$\sigma_{\Delta}=0.34$ms and the accumulated phase statistics over
all nodes were confirmed as marginally white. This validates our noise
assumption stated in Assumption 3. The derived value for $\sigma_{\Delta}$
was used during the experimental validation of Proposition 1 and Proposition
2. No other parameter tuning is needed for the proposed model.

\subsection{Measurement and Simulation Setup}

In both \textsc{Desync} and PCO-based desynchronization, once all
nodes were activated to transmit and receive on a single channel,
a special ``mix message\textquotedblright{} was broadcast by one
of the nodes (chosen randomly) in order to trigger all nodes to set
their initial fire message phase to a random interval within $T=1$s
from its reception. This creates the initial conditions of Assumption
2. The nodes will then desynchronize their transmission of fire messages
and converge to fair TDMA scheduling. We present results under two
convergence thresholds, i.e., $b_{{\text{thres}}}=0.001$ and $b_{{\text{thres}}}=0.020$,
with coupling constants $\alpha\in\left\{ 0.05,\ldots,0.95\right\} $
and number of nodes $W\in\{4,8,16\}$. We use $c_{{\text{conf}}}=0.9999$
to detect convergence under Definition 1 with near certainty. Finally,
we have experimented with various settings for the firing period and
all such experiments led to very similar results for the convergence
iterations to the SState. Thus, all reported experiments use firing
period of $T=1$s, which complies with previous work \cite{ashkiani2012discrete,lien2012anchored,degesys2008towards,Degesys2007DESYNC,scaglione2010bioinspired}.

Under this setup, each node detects convergence to SState by checking
if (\ref{eq:check convergence}) is valid for its last ten firing
cycles. After achieving SState and remaining in this state for 50
firing cycles, a node broadcasts another mix message, in order to
repeat the process. Each node reported the number of its firing cycles
until convergence was detected (minus nine cycles) via a special ``report\textquotedblright{}
message to a base station listening passively to all messages for
monitoring purposes. This facilitated the automated collection of
50 such results per number of nodes, threshold and coupling constant.

In order to cross-validate our theoretical and experimental results
with simulations, we used the Matlab code of Degesys \emph{et al.}
\cite{Degesys2007DESYNC} for \textsc{Desync} and added to it Matlab
code for PCO with inhibitory coupling. In order to simulate the noise
conditions observed in our experimental setup, we deliberately apply
zero-mean additive noise in each phase update with $\sigma_{\Delta}=0.34$ms
and set each node to misfire with probability 0.4\%. Despite the fact
that the simulation cannot capture the complex behavior of the real
system in full detail, it allows for numerous desynchronization processes
to be simulated (300 Matlab runs per triplet $\left\{ W,\alpha,b_{\text{thres}}\right\} $
for each algorithm).

\subsection{\textsc{Desync} Results}

\begin{figure*}
\centering\subfigure[\textsc{Desync}, $W=4$, $b_{\text{thres}}=0.001$]{\includegraphics[width=1\columnwidth]{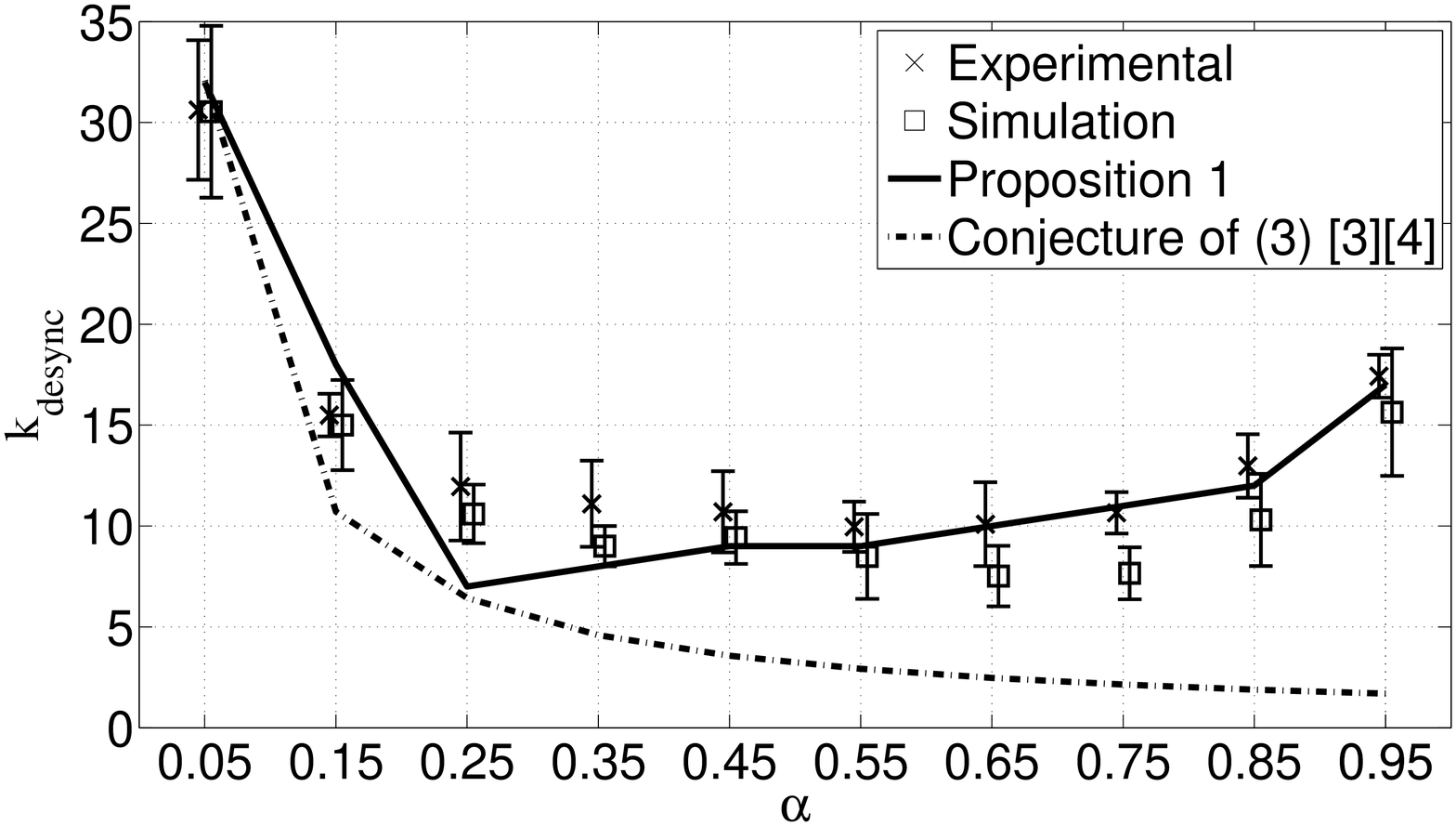}}\centering\subfigure[\textsc{Desync}, $W=4$, $b_{\text{thres}}=0.020$]{\includegraphics[width=1\columnwidth]{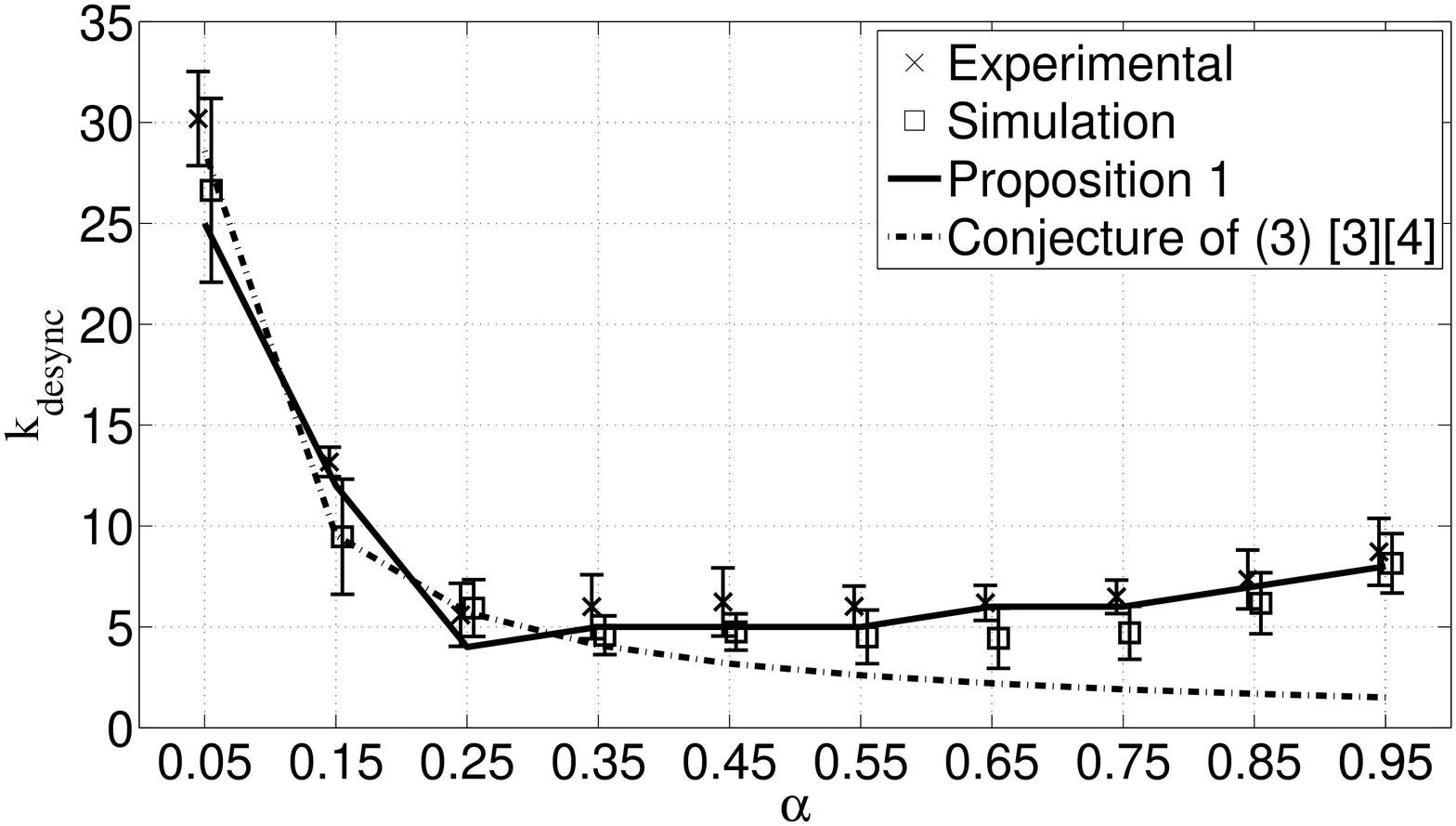}}

\subfigure[\textsc{Desync}, $W=8$, $b_{\text{thres}}=0.001$]{\includegraphics[width=1\columnwidth]{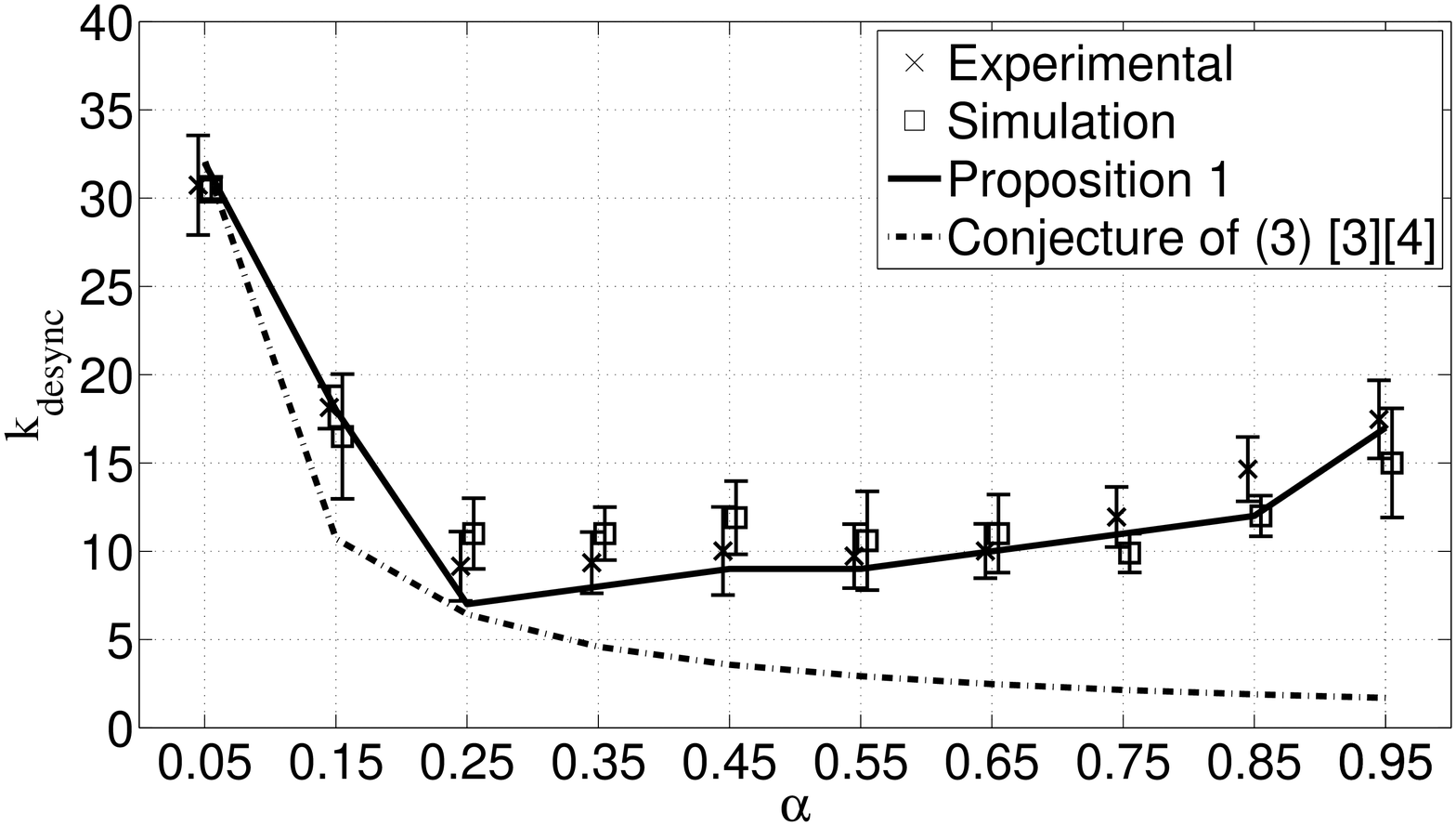}}\subfigure[\textsc{Desync}, $W=8$, $b_{\text{thres}}=0.020$]{\includegraphics[width=1\columnwidth]{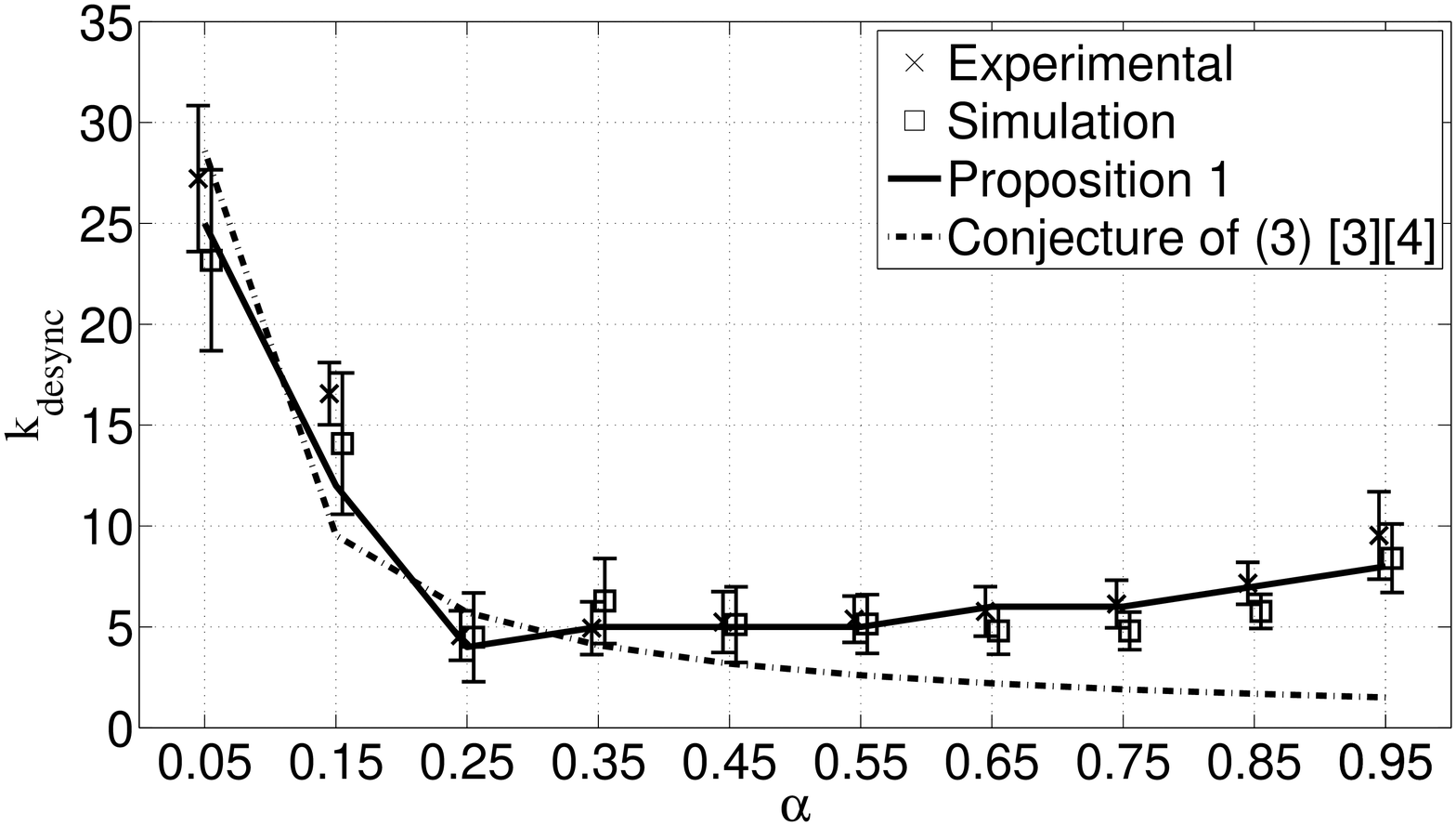}}

\subfigure[\textsc{Desync}, $W=16$, $b_{\text{thres}}=0.001$]{\includegraphics[width=1\columnwidth]{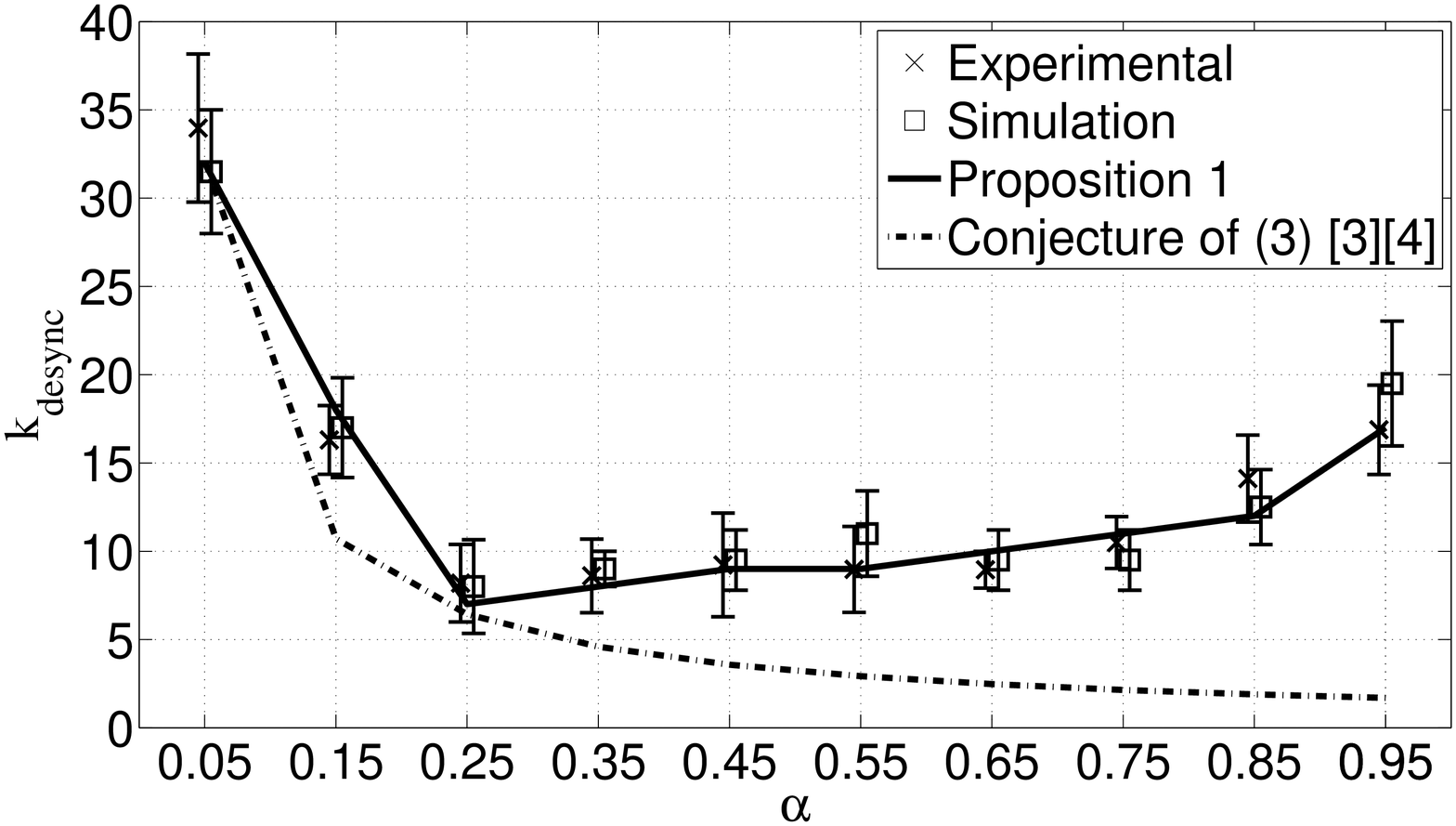}}\subfigure[\textsc{Desync}, $W=16$, $b_{\text{thres}}=0.020$]{\includegraphics[width=1\columnwidth]{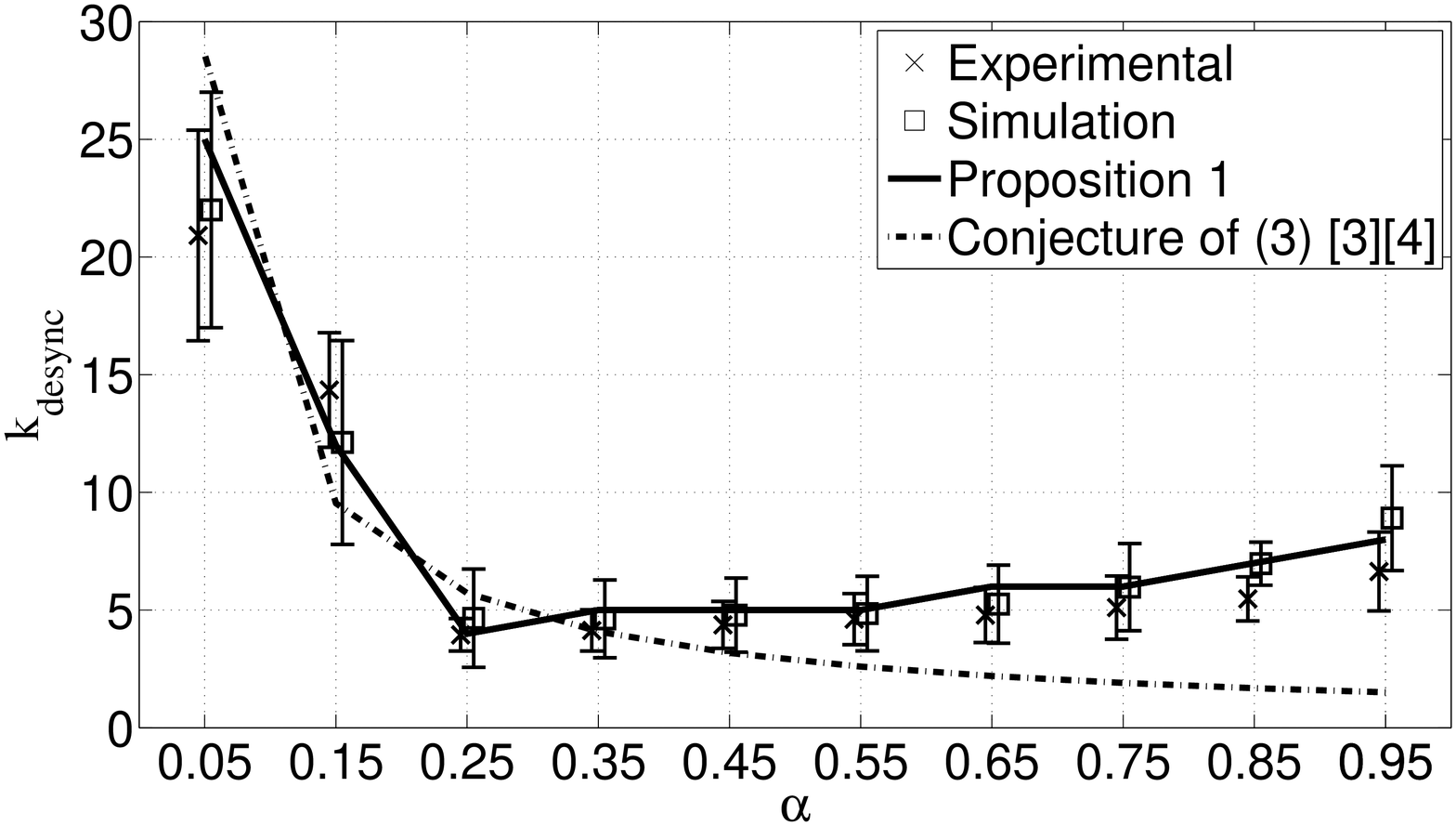}}

\protect\caption{Required firing cycles for convergence to SState for the \textsc{Desync}
algorithm for various values of $\alpha$. The vertical error bars
correspond to one standard deviation from the experimental (or simulation)
mean values, which are indicated by marks. \label{fig:Required-firing-cycles for DESYNC}}
\end{figure*}

The results for this desynchronization mechanism are reported in Fig.
\ref{fig:Required-firing-cycles for DESYNC}. All measurements around
a value of $\alpha$ correspond to results with that value of $\alpha$;
they have been plotted slightly separately solely for ease of illustration.
For comparison purposes, we have also included the order--of--convergence
conjecture of (\ref{eq:conjecture of DESYNC TDMA}) \cite{scaglione2010bioinspired}\cite{Degesys2007DESYNC}
in our results, by scaling the order estimate to fit within the range
of the obtained experiments and simulations. 

The results of Fig. \ref{fig:Required-firing-cycles for DESYNC} show
that the WSN tends to converge to steady state faster when $\alpha$
decreases (until $\alpha=0.25$), since the presence of measurement
noise causes higher-amplitude perturbations for strong coupling, i.e.,
for high values of $\alpha$. However, for very small values of $\alpha$,
the convergence iterations increase dramatically due to weakened coupling
between phase-neighboring nodes. Moreover, by comparing the convergence
results for low and high convergence threshold $b_{{\text{thres}}}$,
one can observe that the use of small convergence threshold increases
the required convergence iterations to SState. 

The proposed model predicts these trends in the convergence iterations
accurately. Specifically, the estimate of Proposition 1 is within
one standard deviation of the experimental and simulation results
for the vast majority of cases. The Pearson correlation coefficients
for the proposed model curves against the mean experimental values
(averaged over $W\in\left\{ 4,\,8,\,16\right\} $) were found to be
$0.9893$ and $0.9931$ for $b_{{\text{thres}}}=0.001$ and $b_{{\text{thres}}}=0.020$,
respectively. The corresponding Pearson correlation coefficients for
the conjecture of \eqref{eq:conjecture of DESYNC TDMA} from \cite{Degesys2007DESYNC,scaglione2010bioinspired}
were $0.8639$ and $0.9464$. These results underline the superior
estimation accuracy of our approach. Finally, the experimental results
of Fig. \ref{fig:Required-firing-cycles for DESYNC} show no statistical
dependence on $W$, which agrees with Proposition 1.

\begin{figure*}
\centering\subfigure[PCO, $W=4$, $b_{\text{thres}}=0.001$]{\includegraphics[width=1\columnwidth]{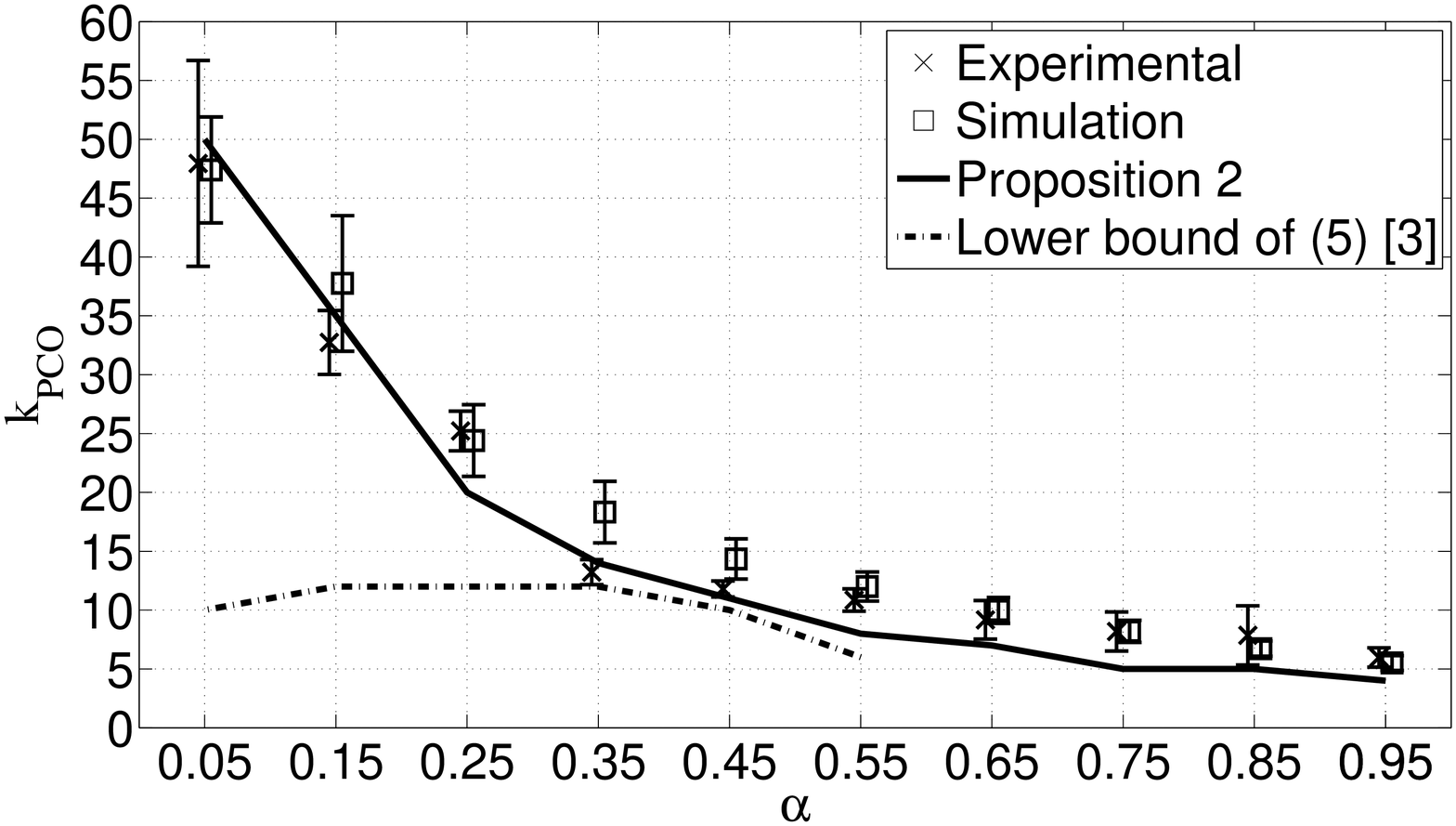}}\centering\subfigure[PCO, $W=4$, $b_{\text{thres}}=0.020$]{\includegraphics[width=1\columnwidth]{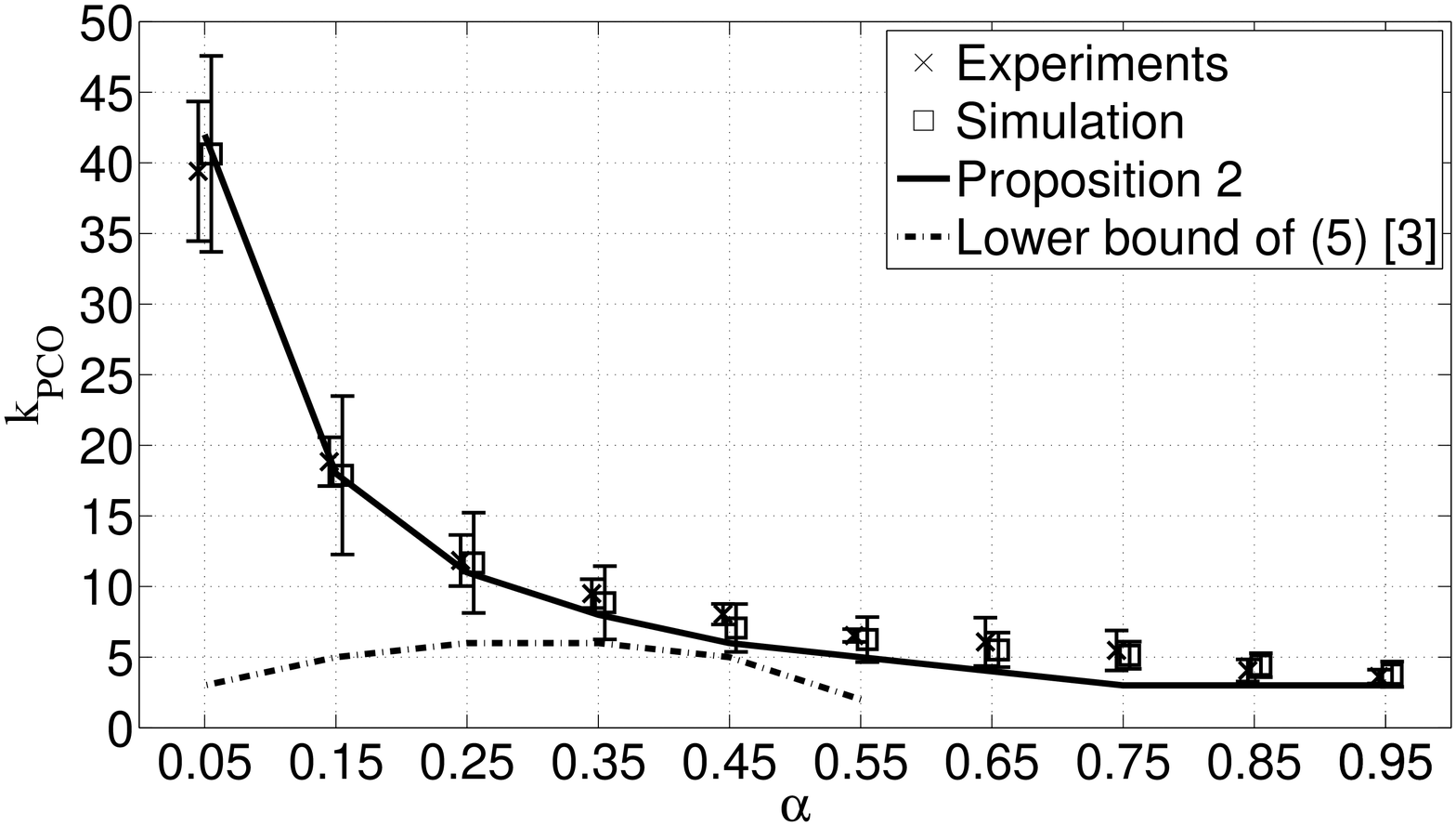}}

\subfigure[PCO, $W=8$, $b_{\text{thres}}=0.001$]{\includegraphics[width=1\columnwidth]{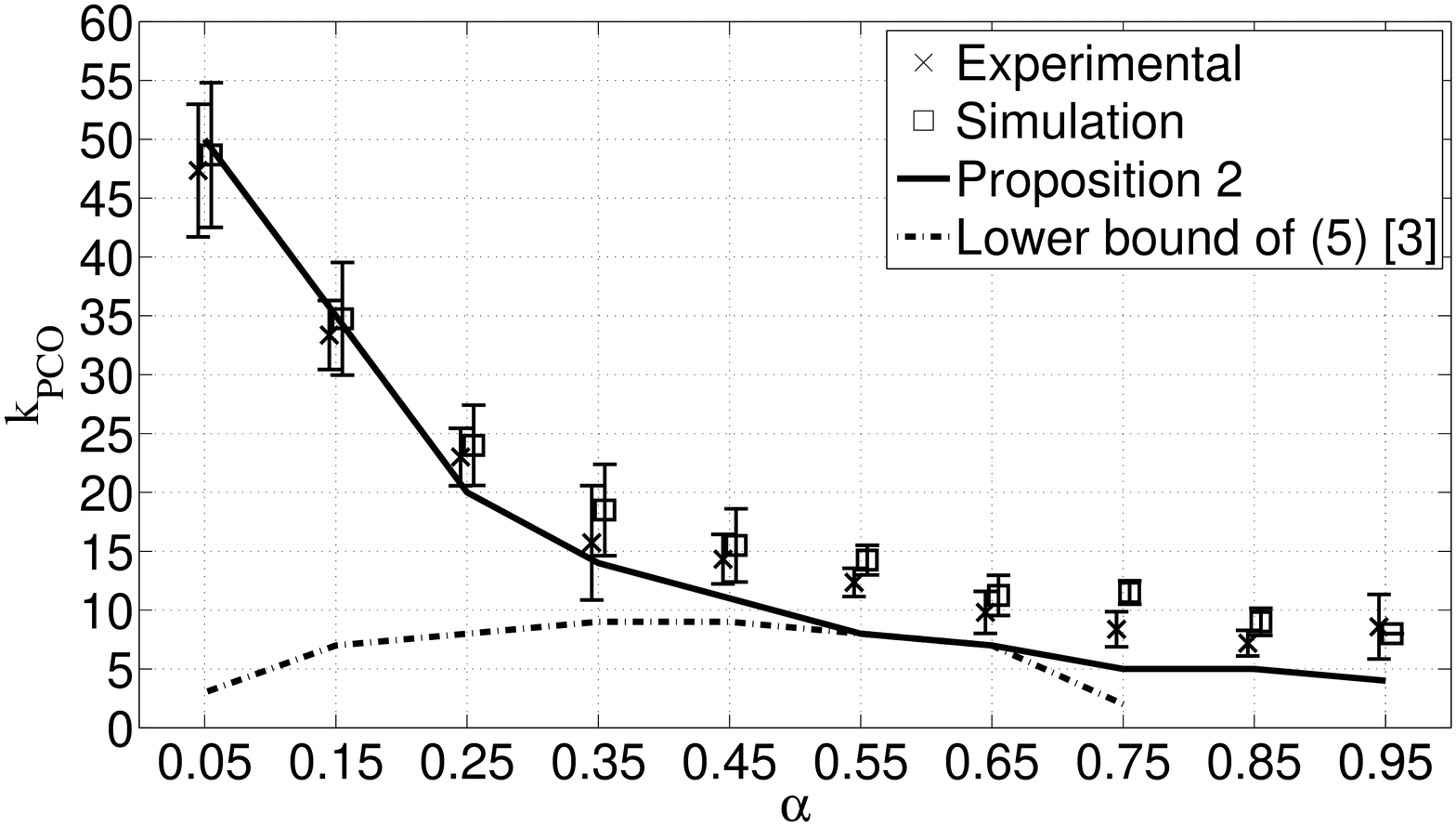}}\subfigure[PCO, $W=8$, $b_{\text{thres}}=0.020$]{\includegraphics[width=1\columnwidth]{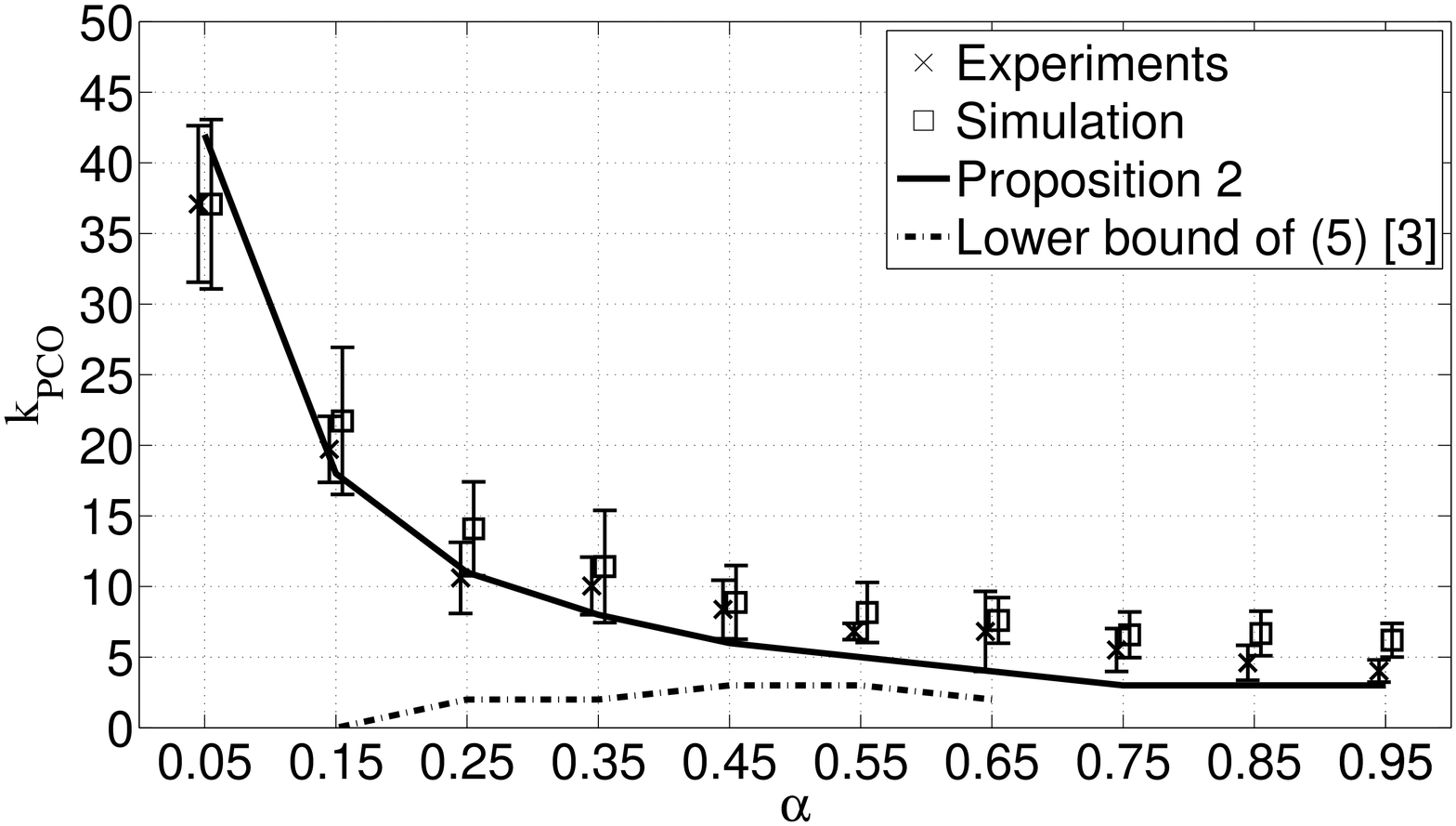}}

\subfigure[PCO, $W=16$, $b_{\text{thres}}=0.001$]{\includegraphics[width=1\columnwidth]{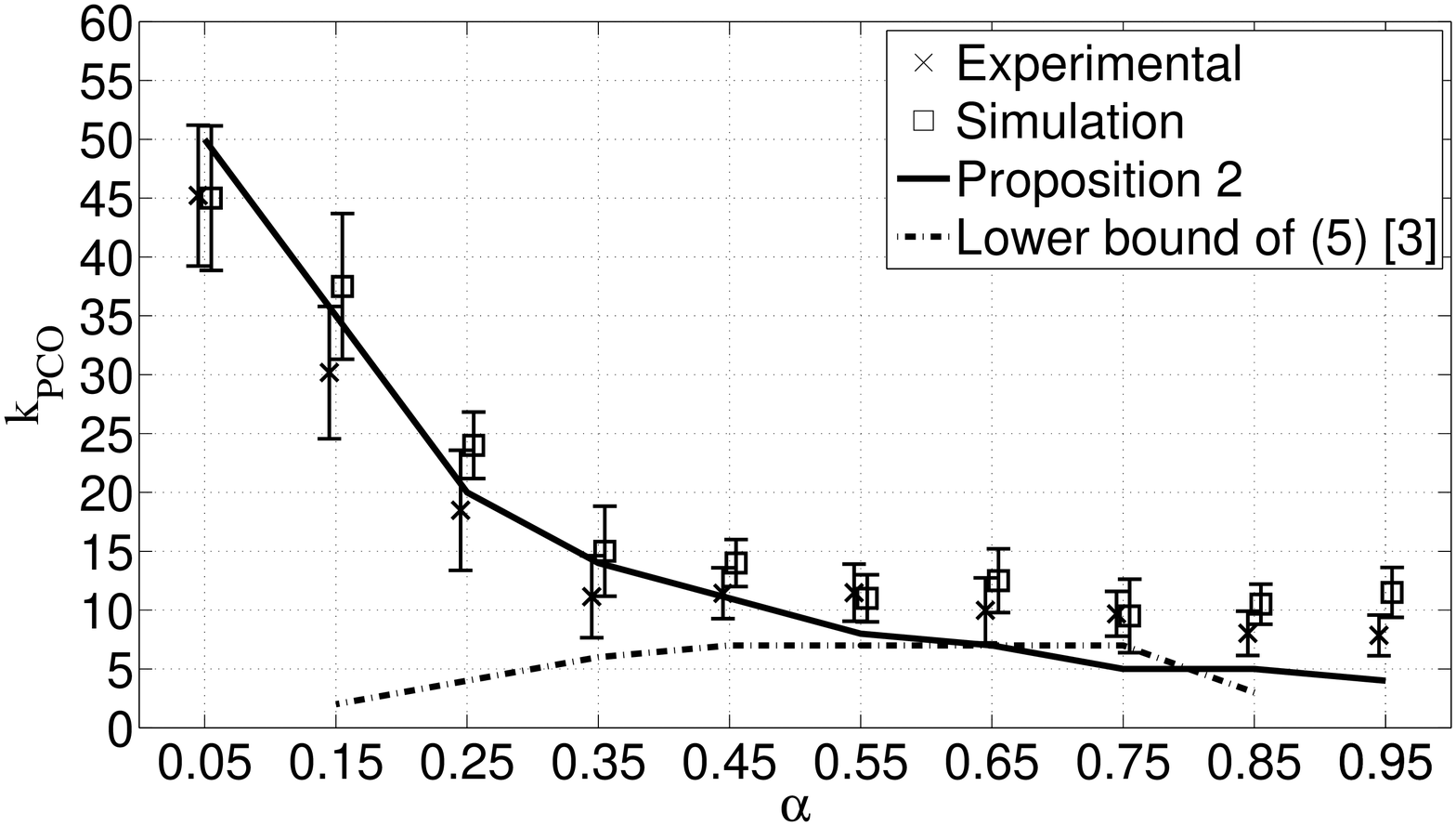}}\subfigure[PCO, $W=16$, $b_{\text{thres}}=0.020$]{\includegraphics[width=1\columnwidth]{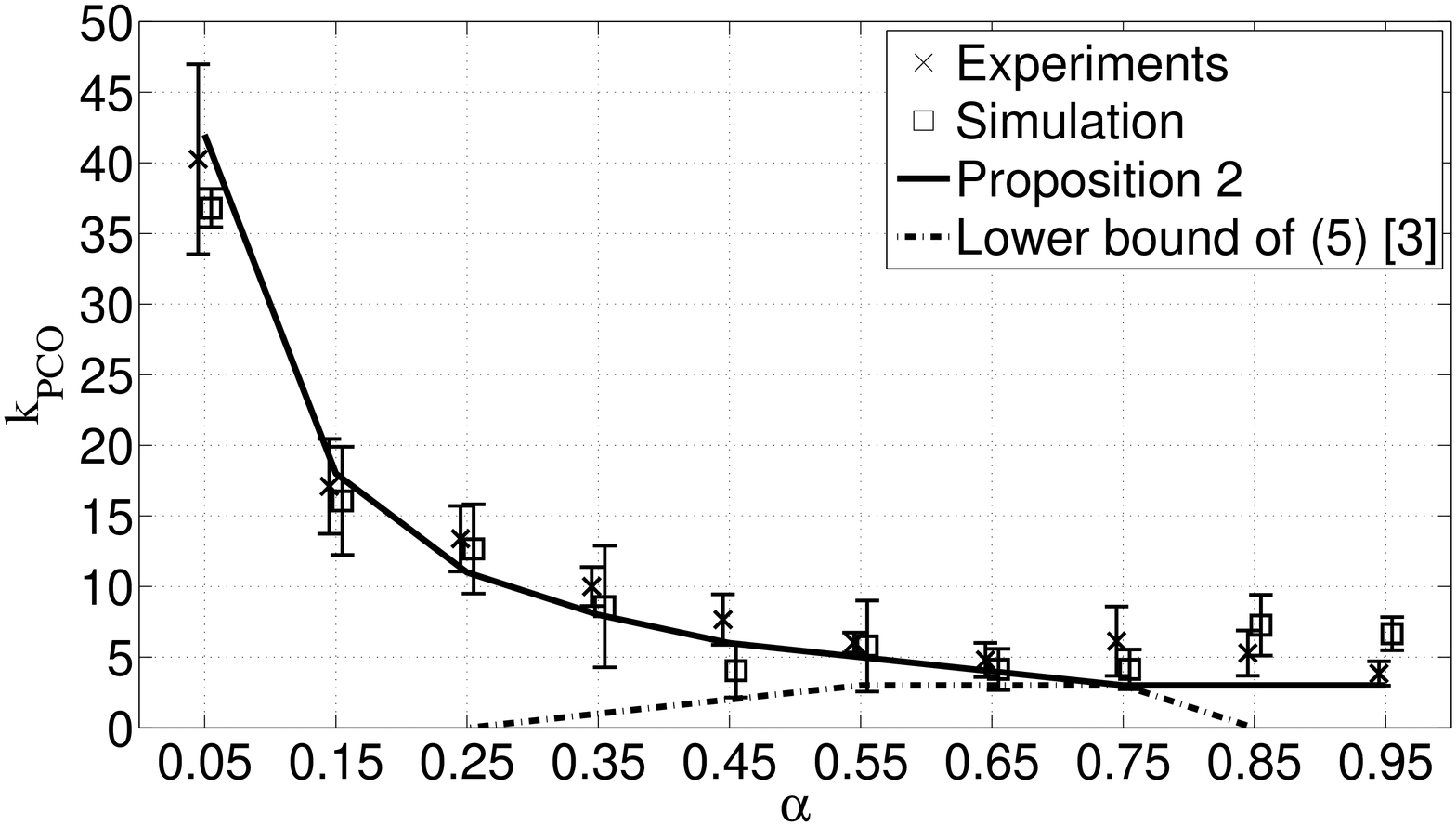}}

\protect\caption{Required firing cycles for convergence to SState for the PCO-based
algorithm for various values of $\alpha$. The vertical error bars
correspond to one standard deviation from the experimental (or simulation)
mean values, which are indicated by marks. \label{fig:Required-firing-cycles for PCO}}
\end{figure*}

\subsection{Results with PCO-based Desynchronization}

The results are reported in Fig. \ref{fig:Required-firing-cycles for PCO}
for both small and large convergence thresholds. Since (\ref{eq:conjecture of PCO})
provided negative estimates for most values of $\alpha$, we added
an offset to the results of the bound to bring as many as possible
to the non-negative region and present only the non-negative ones.
Evidently, the bound of (\ref{eq:conjecture of PCO}) does not match
the observed behavior. We remark however that this is to be expected,
as the bound of (\ref{eq:conjecture of PCO}) is derived under the
assumption that each firing is influenced only by the firing of one
phase-neighboring node \cite{scaglione2010bioinspired}.

Fig. \ref{fig:Required-firing-cycles for PCO} shows that, in PCO,
the convergence iterations decrease monotonically with $\alpha$.
The proposed model predicts this trend correctly and, remains within
one standard deviation from the majority of the experimental and simulation
results. The model results do not change%
\footnote{i.e. Proposition 2 leads to the same results for the convergence iterations
under different values of $W$ %
} for different values of $W$, which agrees with the overall experimentally-observed
behavior of the system. Finally, by comparing the convergence results
for low and high convergence threshold, we note that the use of small
convergence threshold increases the required convergence iterations
to SState. The proposed model predicts this behavior correctly and
agrees with the experimental trends reported. The Pearson correlation
coefficients for the model curves against the mean experimental values
were $0.9739$ for $b_{{\text{thres}}}=0.001$ and $0.9989$ for $b_{{\text{thres}}}=0.020$.

\subsection{Discussion}

By cross referencing between Fig. \ref{fig:Required-firing-cycles for DESYNC}
and Fig. \ref{fig:Required-firing-cycles for PCO} we can compare
the convergence iterations of both algorithms under different settings.
Under appropriate choice for the coupling coefficient, $\alpha$,
the required firing cycles for convergence with PCO-based desynchronization
is comparable to those of \textsc{Desync}. In addition, we can make
the following observations. 
\begin{itemize}
\item \emph{Accuracy of previous analytic work on estimation of convergence
iterations}: Previous estimates or bounds must be scaled to fit the
range of the experimental values, as they are either order--of--convergence
estimates {[}i.e., \eqref{eq:conjecture of DESYNC TDMA}{]}, or overly
optimistic bounds {[}i.e., \eqref{eq:conjecture of PCO}{]}. Moreover,
as illustrated in Fig. \ref{fig:Required-firing-cycles for DESYNC}
and Fig. \ref{fig:Required-firing-cycles for PCO}, when varying the
coupling parameter $\alpha$, previous estimates do not accurately
match the behavior observed in the experimentally-obtained convergence
iterations.
\item \emph{Impact of measurement noise}: Contrary to the proposed model,
these previous estimates do not take into account the measurement
noise conditions. Independently from this work, and following different
analysis and modeling approaches, recent work in synchronization \cite{klinglmayr2012guaranteeing,nishimura2012probabilistic}
and desynchronization \cite{ashkiani2012discrete} has shown that
noise in the desynchronization phase update (e.g., by quantization)
and drops (or collisions) of beacon messages can affect the convergence
iterations and can lead to convergence iterations that deviate from
the estimates obtained based on the ideal (noise-free) model assumed
by earlier work. 
\item \emph{Generalization to non-uniformly distributed initial firings}:
In Propositions 1 and 2, every node's initial phase random variable
was assumed to be i.i.d. uniform via Assumption 2. However, Propositions
1 and 2 hold for any i.i.d. random variable $\Phi^{(1)}$ that satisfies
the three conditions for the generalized form of the central limit
theorem to be applicable \cite[pp. 219-220]{papoulis1989probability}
(see Appendix \ref{sec:Appendix-I}).
\item \emph{Utilized desynchronization algorithm and the network topology}:
While we focused on the initial desynchronization proposals with limited
listening, recent work has extended these initial paradigms towards
other frameworks, where nodes adjust their own firing phase based
on larger listening intervals and/or via the usage of anchored (i.e.,
non-adjusting) nodes \cite{ashkiani2012discrete,choochaisriArtificialForceField,de2012localized,lien2012anchored,motskin2009lightweight,muhlberger2013analyzing,settawatcharawanit2012v}.
While we do not investigate the applicability of our modeling approach
to all such frameworks, our analysis includes the coupling parameter
$\alpha$ and adjusts according to the node firings received within
the predetermined listening interval. It is also worth noting that
Propositions 1 and 2 cover the important scenario of a fully-meshed
(all-to-all) WSN. Desynchronization extensions to multihop scenarios
have been investigated by several works \cite{kang2009localized,motskin2009lightweight,degesys2008towards}.
Given that there is a large variation in the topologies to consider
and that it has been shown that the steady-state difference between
the phases of consecutive firings, as well as the average number of
convergence iterations, varies depending on the network topology under
consideration \cite{nishimura2012probabilistic,motskin2009lightweight,muhlberger2013analyzing,kang2009localized}
(due to the ``selective'' listening of beacons stemming solely from
non-hidden nodes), we do not investigate the validity of our analysis
to such cases. However, we remark that this could be attempted following
the same approach as for Propositions 1 and 2, if the topology specification
is known \emph{a-priori}. 
\item \emph{Consideration of phase updates performed} \emph{during each
node's firing cycle}: As mentioned in the proof of Proposition 1,
to calculate $\Phi_{\text{own}}^{(k)}$ and its moments for \textsc{Desync}
and PCO-based convergence, our analysis does not take into account
the variability in the neighboring nodes' phase updates%
\footnote{e.g., updates that may have been carried out during the $k$th firing
cycle of the node under consideration, or variability due to misfiring
or other non-idealities not captured by our noise assumption (Assumption
2)%
}. This is in agreement with the way \textsc{Desync} is applied in
practice and constitutes a simplification for PCO-based desynchronization
\cite{Degesys2007DESYNC,scaglione2010bioinspired}. Given that our
stochastic modeling framework is in good agreement with the average
experimental and simulation results without requiring experimental
tuning (besides knowledge of the standard deviation of the phase measurement
noise), our approach forms an important step towards considering stochastic
models for the convergence of desynchronization systems. 
\end{itemize}

\section{Application Examples}

The proposed stochastic estimation framework can benefit desynchronization-based
TDMA protocols in WSNs by analytically estimating the impact of the
phase-coupling constant $\alpha$, convergence threshold $b_{{\text{thres}}}$
and firing cycle period $T$ on such deployments \cite{scaglione2010bioinspired,Degesys2007DESYNC,buranpanichkit2012distributed}.
We present two such examples.

\subsection{Maximizing the Bandwidth Per Node }

In the first case, we consider a WSN that initially comprises $W$
nodes, where nodes are expected to join or leave the network every
$T_{\text{swap}}$ seconds. This scenario usually occurs when mobile
nodes periodically enter and exit the coverage area of the network,
or some nodes switch on and off periodically to conserve energy. In
fair TDMA scheduling, the bandwidth per node is $\frac{B_{\text{WSN}}}{W}$
bps, where $B_{\text{WSN}}$ is the maximum application-layer bandwidth%
\footnote{Following the approach of Degesys \cite{Degesys2007DESYNC}, we estimate
$B_{\text{WSN}}$ by using a single transmitter and receiver to measure
the achieved delivery rate at the application layer.%
} in IEEE 802.15.4. In practice, the fluctuating number of nodes in
the WSN will result in bandwidth loss as, each time nodes join or
leave, the system needs to converge to SState before transmission
resumes with equal slot size \cite{buranpanichkit2012distributed,Degesys2007DESYNC,BesbesTWC2013}.
Using the proposed framework, we can derive an estimate of the expected
bandwidth per node under such conditions. Specifically, if the PCO
or \textsc{Desync} firing-cycle period is $T$ seconds and the node
joining or exiting occurs (on average) every $T_{\text{swap}}$ seconds,
the expected bandwidth per node can be estimated as
\begin{equation}
B_{\text{swap}}=\left(1-\frac{k_{\text{method}}T}{T_{\text{swap}}}\right)\frac{B_{\text{WSN}}}{W},\;\text{method}\in\left\{ \text{desync, PCO}\right\} ,\label{eq:B_loss}
\end{equation}
with $k_{\text{method}}$ the expected firing cycles until convergence
to SState, given by Proposition 1 or Proposition 2. The factor $\frac{k_{\text{method}}T}{T_{\text{swap}}}$
in (\ref{eq:B_loss}) expresses the normalized loss of bandwidth per
node due to convergence to TDMA under \textsc{Desync} and PCO every
time nodes join or leave the network. 

\begin{table*}
\noindent \begin{centering}
\protect\caption{\label{tab:Desync Throughput}\textsc{Desync}: average bandwidth per
node, $B_{\text{\text{swap}}}$ (in kbps), under different convergence
thresholds and different values for the phase-coupling constant $\alpha$.}

\par\end{centering}

\noindent \centering{}%
\begin{tabular}{|c|c|c|c|c|}
\hline 
\textsc{Desync} & \multicolumn{2}{c|}{$b_{{\text{thres}}}=0.001$ (1ms) } & \multicolumn{2}{c|}{$b_{{\text{thres}}}=0.020$ (20ms)}\tabularnewline
\hline 
Phase-coupling constant & Measurement & Theoretical via (\ref{eq:B_loss}) and Prop. 1 & Measurement & Theoretical via (\ref{eq:B_loss}) and Prop. 1\tabularnewline
\hline 
\hline 
$\alpha=0.95$ \cite{Degesys2007DESYNC} & 6.55 & 7.14 & 7.72 & 7.91\tabularnewline
\hline 
$\alpha=0.25$ (from Fig. \ref{fig:Required-firing-cycles for DESYNC}) & 7.41 & 8.00 & 8.17 & 8.26\tabularnewline
\hline 
Average bandwidth gain per node ($\%$)  & 13.13 & 12.04 & 5.83 & 4.42\tabularnewline
\hline 
\end{tabular}
\end{table*}

\begin{table*}
\noindent \begin{centering}
\protect\caption{\label{tab:PCO Throughput}PCO: average bandwidth per node, $B_{\text{\text{swap}}}$
(in kbps), under different convergence thresholds and different values
for the phase-coupling constant$\alpha$.}

\par\end{centering}

\noindent \centering{}%
\begin{tabular}{|c|c|c|c|c|}
\hline 
\textsc{PCO} & \multicolumn{2}{c|}{$b_{{\text{thres}}}=0.001$ (1ms) } & \multicolumn{2}{c|}{$b_{{\text{thres}}}=0.020$ (20ms)}\tabularnewline
\hline 
Phase-coupling constant & Measurement & Theoretical via (\ref{eq:B_loss}) and Prop. 2 & Measurement & Theoretical via (\ref{eq:B_loss}) and Prop. 2\tabularnewline
\hline 
\hline 
$\alpha=0.75$ \cite{scaglione2010bioinspired} & 6.90 & 8.17 & 7.65 & 8.34\tabularnewline
\hline 
$\alpha=0.95$ (from Fig. \ref{fig:Required-firing-cycles for PCO}) & 7.54 & 8.26 & 8.13 & 8.34\tabularnewline
\hline 
Average bandwidth gain per node ($\%$)  & 9.28 & 1.10 & 6.27 & 0.00\tabularnewline
\hline 
\end{tabular}
\end{table*}

Using the experimental setup of Section IV, we present an example
of the calculation of (\ref{eq:B_loss}) for a WSN comprising $W=10$
nodes with $T=1$s and $1\sim3$ nodes entering or exiting the network
every $T_{\text{swap}}$ seconds, with $T_{\text{swap}}\in\left[70,130\right]$
to incorporate up to 30$\%$ variability around the mean value of
$100$s. The maximum application-layer bandwidth was measured to be
$\textsc{ \ensuremath{B_{\text{WSN}}=86}}$kbps. Table \ref{tab:Desync Throughput}
and Table \ref{tab:PCO Throughput} present the results when using
the value of $\alpha$ that was estimated to provide the minimum $k_{\text{method}}$
under Propositions 1 and 2 against the result when using values for
$\alpha$ suggested from previous work. It is evident that, for all
cases, the proposed model provides the setting for $\alpha$ that
minimizes the convergence iterations and leads to the maximum achievable
bandwidth per node. This is despite the fact that the model assumes
uniformly-distributed initial fire times, while in the application
results the values of most nodes may be (approximately) equidistant
at the moment $1\sim3$ nodes enter or exit the network. The impact
on the achieved bandwidth per node is much more pronounced in the
case of \textsc{Desync}, where previous work used $\alpha=0.95$ instead
of the best option, which is $\alpha=0.25$. The results in Table
\ref{tab:Desync Throughput} and Table \ref{tab:PCO Throughput},
show that selecting $\alpha$ through the proposed model brings gains
of up to 13$\%$ in the bandwidth per node compared with the standard
settings used in existing works \cite{Degesys2007DESYNC,scaglione2010bioinspired}.

\subsection{Estimating the Required Firing Cycle Period }

In the second application example, we focus on the case of a WSN using
\textsc{Desync} or PCO and requiring convergence to steady state to
be achieved within $T_{\text{SState}}$ seconds on average. The desired
value of $T_{\text{SState}}$ depends on the application context,
e.g., a predefined value $T_{\text{SState}}$ in order to limit the
delay and buffering requirements between the data acquisition and
transmission when the WSN is activated \cite{bojic2012self,buranpanichkit2012distributed,BesbesTWC2013}. 

Based on Proposition 1 and Proposition 2, and under given settings
for $b_{{\text{thres}}}$ and $\alpha$, we can match the period of
firing cycles ($T$) to the desired convergence time by: $k_{\text{method}}T=T_{\text{SState}}$.
By solving the last equation for $T$ we derive the firing cycle period
that meets the convergence time expectation under $k_{\text{method}}$
given by Proposition 1 and Proposition 2 for each setting of $b_{{\text{thres}}}$
and $\alpha$. 

\begin{table*}
\noindent \begin{centering}
\protect\caption{\label{tab:Convergence time results}\textsc{Desync} and PCO for desired
convergence time of $10$s. The corresponding firing cycle period,
$T$, is given in parentheses (in seconds) in the ``Theoretical''
columns.}

\par\end{centering}

\noindent \centering{}%
\begin{tabular}{|c|c|c|c|c|}
\hline 
\multirow{2}{*}{Method} & \multicolumn{2}{c|}{\textsc{Desync}} & \multicolumn{2}{c|}{PCO}\tabularnewline
\cline{2-5} 
 & Experimental $T_{\text{SState}}$ & Theoretical $T_{\text{SState}}$ & Experimental $T_{\text{SState}}$ & Theoretical $T_{\text{SState}}$\tabularnewline
\hline 
\hline 
$\alpha=0.25$$\textsc{,}$ $b_{{\text{thres}}}=0.001$ & 14.01 & 10.00 (1.43) & 7.80 & 10.00 (0.50)\tabularnewline
\hline 
$\alpha=0.95$$\textsc{,}$ $b_{{\text{thres}}}=0.001$ & 12.15 & 10.00 (0.59) & 18.75 & 10.00 (2.50)\tabularnewline
\hline 
$\alpha=0.25$$\textsc{,}$ $b_{{\text{thres}}}=0.020$ & 11.75 & 10.00 (2.50) & 10.83 & 10.00 (0.91)\tabularnewline
\hline 
$\alpha=0.95$$\textsc{,}$ $b_{{\text{thres}}}=0.020$ & 10.38 & 10.00 (1.25) & 12.65 & 10.00 (3.33)\tabularnewline
\hline 
\end{tabular}
\end{table*}

We have experimented with various parameters of \textsc{Desync} or
PCO and the obtained theoretical and experimental results are reported
in Table \ref{tab:Convergence time results} using the average obtained
from multiple runs with $W\in\left\{ 4,8,16\right\} $. With the exception
of two cases (\textsc{Desync} at: $\alpha=0.25$$\textsc{,}$ $b_{{\text{thres}}}=0.001$
and PCO at: $\alpha=0.95$$\textsc{,}$ $b_{{\text{thres}}}=0.001$),
the theoretical prediction is within a 25$\%$ margin of the experimentally
observed convergence time.

\section{Conclusions}

A novel stochastic estimation framework for the convergence iterations
to fair TDMA scheduling is proposed for the two desynchronization
primitives with limited listening, namely, \textsc{Desync} and pulse-coupled
oscillators with inhibitory coupling. Our stochastic estimates establish
the expected firing cycles until each node's firing converges to the
steady state with very high confidence. For both algorithms, our analytic
expressions are validated based on simulations and experiments with
a fully-meshed network of wireless sensors. The results show that
our estimates are more accurate than previous order-of-convergence
estimates and lower bounds. Our model incorporates the influence of
system parameters (i.e., total number of nodes, coupling coefficient,
convergence threshold and phase measurement noise) on the expected
convergence iterations. Therefore, it can be used to estimate the
best operational parameters (and the associated delay) to establish
fair TDMA scheduling under several desynchronization-based WSN protocols.
More broadly, the analysis of this paper contributes towards the analytic
understanding of how a desynchronization system is expected to evolve
from random initial conditions to the desynchronized steady state.

\appendices 

\section{\label{sec:Appendix-I}}

We show that $\Phi_{\text{own}}^{\left(k\right)}$ of \eqref{eq:k-th iteration of DESYNC update}
and \eqref{eq:PCO phase RV of kth phase update}, i.e., the RV modeling
each node's own phase during its $k$th firing cycle under \textsc{Desync}
and PCO (respectively), is normally distributed after a few phase
updates. 

Once \eqref{eq:k-th iteration of DESYNC update} and \eqref{eq:PCO phase RV of kth phase update}
are reached, we can make the following observations:
\begin{itemize}
\item $\Phi_{\text{own}}^{(k)}$ is a linear mixture of independent random
variables, i.e.: \emph{(i)} i.i.d. noise and phase vectors $\mathbf{\Delta}$
and $\mathbf{\Phi^{(1)}}$ in \textsc{Desync}; \emph{(ii)} $\Phi_{\text{own}}^{\left(1\right)}$
and $\forall l,j:\;\;\Delta_{\text{own}}^{(l-j+1)}$ in PCO;
\item for \textsc{Desync}: $\forall k\in\mathbb{N}^{*}$, we can pick $\varepsilon=(1-\alpha)^{k}\sigma_{\Phi^{(1)}}$
and, from (\ref{eq:std of DESYNC}), $\sigma_{{\text{desync}}}^{(k)}>\varepsilon$
;
\item for PCO: $\forall l\in\mathbb{N}^{*}:$ $\sigma_{{\text{PCO}},l}>(1-\alpha)^{l}\sigma_{\Phi^{(1)}}$;
\item all initial PDFs have finite support (they are all variants of the
uniform distribution); hence, densities ${\text{P}}_{\Phi_{\text{own}}^{(k)}}$
will have finite support since they are linear mixtures of PDFs with
finite support.
\end{itemize}
These observations satisfy the three conditions for the generalized
form of the central limit theorem to be applicable \cite[pp. 219-220]{papoulis1989probability},
and thus $\Phi_{\text{own}}^{\left(k\right)}$ becomes a normally-distributed
random variable after a few phase updates. Papoulis \cite[pp. 219-220]{papoulis1989probability}
suggests that the normal PDF is accurately approximated with just
5 linear combinations of PDFs satisfying the above criteria, meaning
that five phase updates will suffice for the normal distribution to
be an accurate approximation of the firing phase of each node.

\section{\label{sec:Appendix-II}}

We present the analysis that matches the expected number of phase
updates until convergence to the expected number of firing cycles
and concludes the proof of \eqref{eq:firing cycles for PCO convergence}.
\\

\begin{figure*}
\begin{centering}
\includegraphics[scale=0.21]{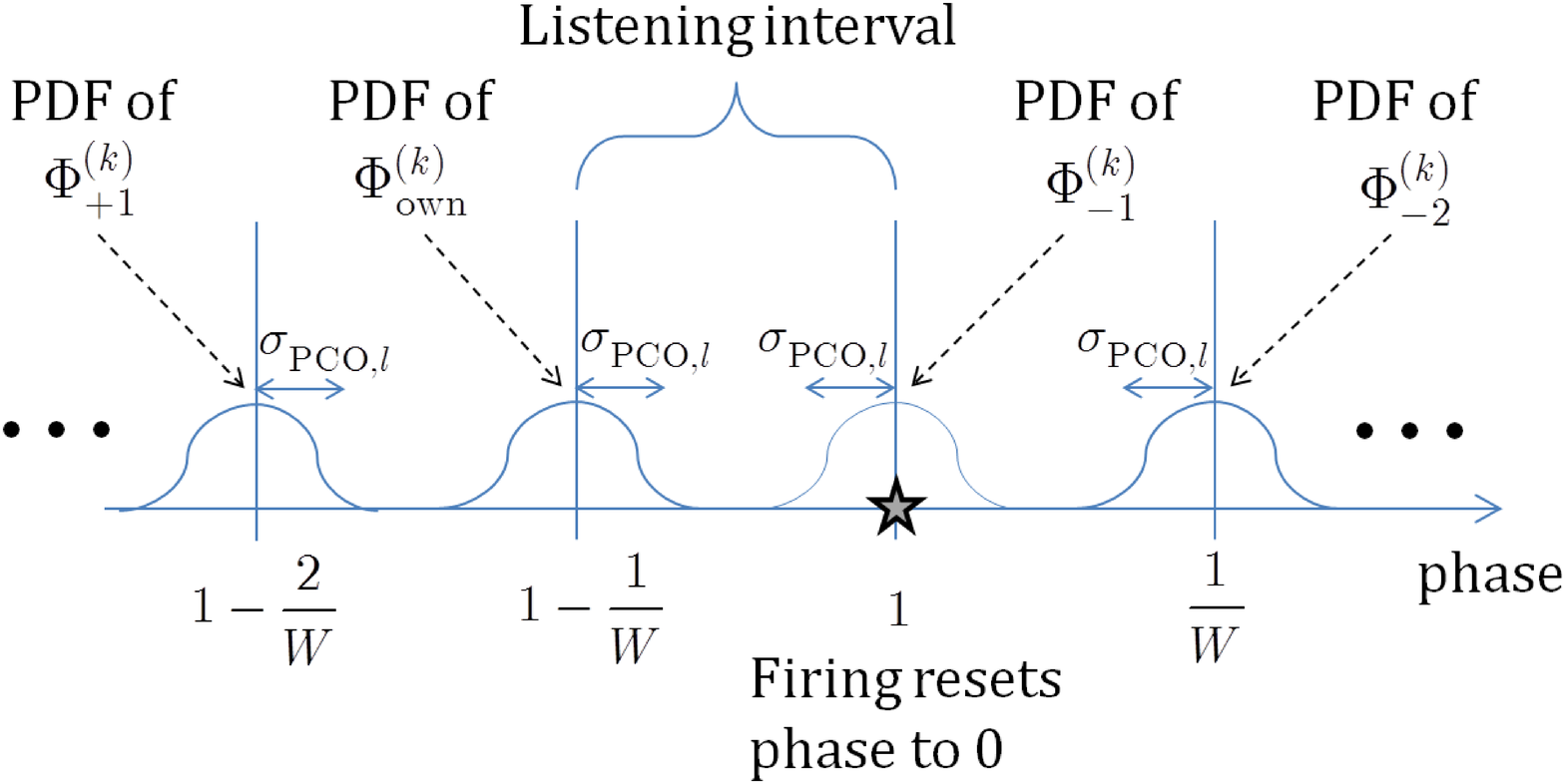} 
\par\end{centering}

\protect\caption{A pictorial illustration of the PDFs of the phase RVs $\left\{ \ldots,\Phi_{-2}^{\left(k\right)},\Phi_{-1}^{\left(k\right)},\Phi_{\text{own}}^{\left(k\right)},\Phi_{+1}^{\left(k\right)},\ldots\right\} $
of the $k$th firing cycle of ``$\text{own}$'' node, during ``$\text{own}$''
node's $l$th phase update, performed via (\ref{eq:PCO phase update}).}

\label{fig:PDF for phase update of PCO} 
\end{figure*}

\textbf{Subsequent firing cycles, effect of phase-neighboring firings:}
A pictorial illustration of the PDF of the $l$th phase update of
a node (performed during its $k$th firing cycle, with $k,l\geq2$)
is given in Fig. \ref{fig:PDF for phase update of PCO} in conjunction
with its listening interval and the PDFs of the two previous firings
and the next firing. Since all phase RVs are normally distributed
after a few phase updates, it is straightforward to infer from Fig.
\ref{fig:PDF for phase update of PCO} that the probability for the
previous firing (represented by RV $\Phi_{-1}^{\left(k\right)}$)
to occur within the listening interval is
\[
\frac{1}{2}\text{erf}\left(\frac{1}{W\sigma_{\text{PCO},l}\sqrt{2}}\right).
\]
Moreover, the probability that $\Phi_{-2}^{\left(k\right)}$ will
occur within the node's listening interval is
\[
\frac{1}{2}\left[\text{erf}\left(\frac{2}{W\sigma_{\text{PCO},l}\sqrt{2}}\right)-\text{erf}\left(\frac{1}{W\sigma_{\text{PCO},l}\sqrt{2}}\right)\right].
\]
 This is also the probability that $\Phi_{+1}^{\left(k\right)}$ will
occur within the node's listening interval.

\textbf{Subsequent firing cycles and the effect of all firings within
a window of $W$ firing events:} We can now generalize the previous
calculation to the probability of occurrence of the $\left\lfloor \frac{W}{2}\right\rfloor $
previous and next firings within the node's listening interval. Beyond
the previous firing, for the $j$th firing after the node's own firing
or the $\left(j+1\right)$th firing before the node's own firing ($1\leq j\leq\left\lfloor \frac{W}{2}\right\rfloor $),
this probability is
\[
\frac{1}{2}\left[\text{erf}\left(\frac{j+1}{W\sigma_{\text{PCO},l}\sqrt{2}}\right)-\text{erf}\left(\frac{j}{W\sigma_{\text{PCO},l}\sqrt{2}}\right)\right].
\]
Hence, summing up the probabilities of firings occurring within the
node's listening interval for all $\left\lfloor \frac{W}{2}\right\rfloor $
firings before and after the current one, the expected number of phase
updates is given by the expression in the summation term of (\ref{eq:firing cycles for PCO convergence}),
where we used the identity ($\forall b\in\mathbb{N}^{*},\forall c\neq0$):
\begin{eqnarray}
\sum_{j=1}^{b}\left[\text{erf}\left(\frac{j+1}{c}\right)-\text{erf}\left(\frac{j}{c}\right)\right]+\frac{1}{2}\text{erf}\left(\frac{1}{c}\right)\nonumber\\
=\text{erf}\left(\frac{b+1}{c}\right)-\frac{1}{2}\text{erf}\left(\frac{1}{c}\right).
\end{eqnarray}

The expected number of phase updates within $k$ firing cycles of
a node is
\[
\sum\limits _{l=2}^{k}\left[\text{erf}\left(\frac{\left\lfloor \frac{W}{2}\right\rfloor +1}{W\sigma_{\text{PCO},l}\sqrt{2}}\right)-\frac{1}{2}\text{erf}\left(\frac{1}{W\sigma_{\text{PCO},l}\sqrt{2}}\right)\right]+1-\frac{1}{W}.
\]
 As a result, for $l_{\text{SSupd}}$ phase updates leading to convergence
under Definition 1 {[}shown by \eqref{eq:number of phase update of PCO}{]},
the corresponding number of firing cycles is given by (\ref{eq:firing cycles for PCO convergence}). 

 \bibliographystyle{IEEEtran}
\bibliography{literature}

\begin{IEEEbiography}[{\includegraphics[width=1in,height=1.25in,clip,keepaspectratio]{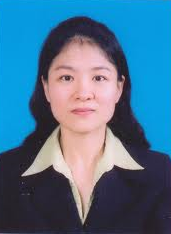}}]{Dujdow Buranapanichkit}
 obtained the PhD from the Department of Electronic and Electrical Engineering of University College London (UK). She is now Lecturer in the Prince of Songkla University, Thailand. 

Her research interests are in wireless sensor networks and distributed synchronization mechanisms and protocol design.
\end{IEEEbiography}

\begin{IEEEbiography}[{\includegraphics[width=1in,height=1.25in,clip,keepaspectratio]{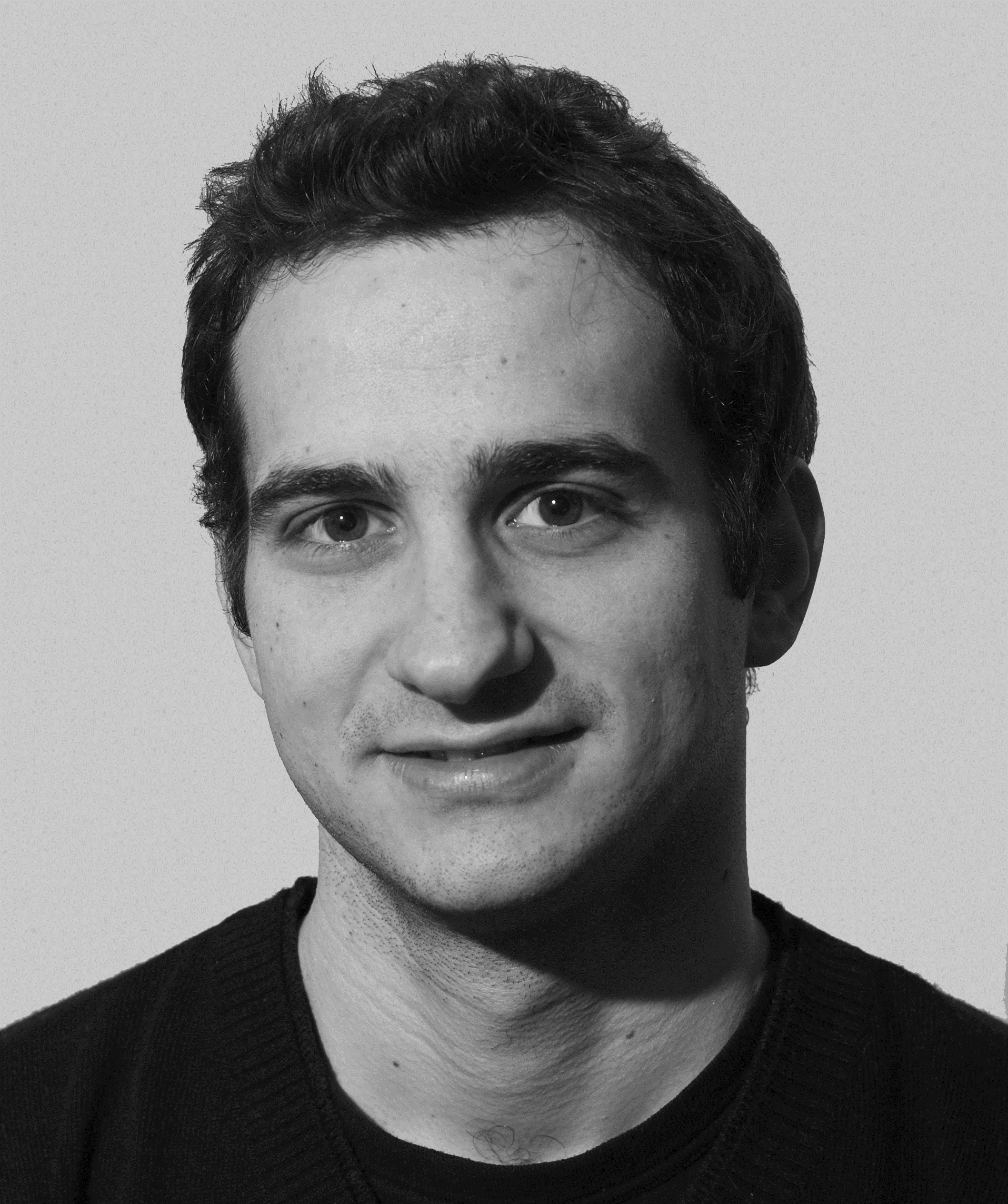}}]{Nikos Deligiannis}
(M\textquoteright10) received the Diploma in electrical and computer engineering from University of Patras, Greece, in 2006, and the PhD in applied sciences (awarded with highest distinction and congratulations from the jury members) from Vrije Universiteit Brussel, Brussels, Belgium, in 2012.
 
From June 2012 to September 2013, he was a Postdoctoral Researcher with the Department of Electronics and Informatics, Vrije Universiteit Brussel, Brussels, Belgium, and a Senior Research Engineer with the 4Media Group of the iMinds research institute, Ghent, Belgium. In October 2013, he joined the Department of Electronic and Electrical Engineering at University College London, where he is currently working as Postdoctoral Research Associate. His research interests include multiterminal communications, multidimensional and sparse signal processing, distributed source coding, wireless sensor networks, and multimedia systems.
 
Dr. Deligiannis has received the 2011 ACM/IEEE International Conference on Distributed Smart Cameras Best Paper Award and the 2013 Scientific Prize IBM-FWO Belgium.
\end{IEEEbiography}

\begin{IEEEbiography}[{\includegraphics[width=1in,height=1.25in,clip,keepaspectratio]{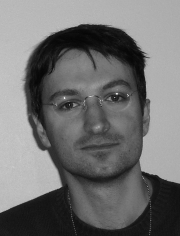}}]{Yiannis
Andreopoulos}
(M\textquoteright00, SM\textquoteright14) is Senior Lecturer in the Department of Electronic and Electrical Engineering of University College London (UK). His research interests are in wireless sensor networks, error-tolerant computing and multimedia systems.

He received the 2007 �Most-Cited Paper� award
from the Elsevier EURASIP Signal Processing:
Image Communication journal and a best-paper
award from the 2009 IEEE Workshop on Signal
Processing Systems. Dr. Andreopoulos was Special
Sessions Co-chair of the 10th International Workshop
on Image Analysis for Multimedia Interactive Services (WIAMIS 2009)
and Programme Co-chair of the 18th International Conference on Multimedia
Modeling (MMM 2012) and the 9th International Conference on Body
Area Networks (BODYNETS 2014). He is an Associate editor of the IEEE
TRANSACTIONS ON MULTIMEDIA, the IEEE SIGNAL PROCESSING LETTERS and
the Elsevier Image and Vision Computing journal.

\end{IEEEbiography}

\end{document}